\documentclass[prd,aps,nofootinbib,superscriptaddress]{revtex4}
\usepackage{epsfig}
\usepackage{amsmath, slashed}
\usepackage{xcolor}
\usepackage{appendix}
\usepackage{dsfont}
 \usepackage{mathrsfs}
\usepackage[normalem]{ulem}

\usepackage{bm}
\usepackage{amsfonts}

\allowdisplaybreaks
\begin{document}

\title{Accessing baryon-antibaryon generalized distribution amplitudes in $e^{\pm} \gamma \to  e^{\pm} B  \bar{B} $}

\author{Jing Han}
\affiliation{School of Physics, Zhengzhou University, Zhengzhou, Henan 450001, China}
\author{Bernard Pire }
\affiliation{CPHT, CNRS, \'Ecole polytechnique, Institut Polytechnique de Paris, 91128 Palaiseau, France}
\author{Qin-Tao Song}
\email[]{songqintao@zzu.edu.cn}
\affiliation{School of Physics, Zhengzhou University, Zhengzhou, Henan 450001, China}

\date{\today}

\begin{abstract}
{$\gamma^* \gamma \to B \bar B$ is the golden process to access chiral-even di-baryon generalized distribution amplitudes (GDAs) as deeply virtual Compton scattering has proven to be for the generalized parton distributions. In the framework of colinear QCD factorization where the leading twist amplitude is the convolution of GDAs and a perturbatively calculable coefficient function, we study the scattering amplitude for the baryonic channels where $B$  is a spin $1/2$ baryon such as the nucleon or hyperon $\Lambda, \Sigma$. Taking into account the interfering QED amplitude, we calculate the cross section of the $e^\pm \gamma \to e^\pm B \bar B$ process that can be experimentally studied in $e^- e^+$ as well as in electron–ion facilities. We explore both the final state polarization summed case and the polarization dependent effects.  Numerical estimates are presented  for $e^- \gamma \to e^- p \bar p$, using motivated  models for  GDAs. Our results show that a first extraction of baryon-antibaryon GDAs from experimental measurements is feasible at Belle II. }
\end{abstract}

\maketitle

\date{}

\section{Introduction}
\label{introduction}
Generalized distribution amplitudes (GDAs)~\cite{Muller:1994ses,Diehl:1998dk,Polyakov:1998ze} and the closely related generalized parton distributions (GPDs)~\cite{Muller:1994ses,Ji:1996ek,Radyushkin:1998bz} are hadronic matrix elements  of light-cone bilocal operators that can be used for a hadron tomography~\cite{Burkardt:2000za,Ralston:2001xs,Diehl:2002he,Pire:2002ut}. Moreover, GDAs and GPDs offer an indirect access to the gravitational form factors (GFFs), which in turn enable us to address the longstanding puzzle of nucleon spin decomposition~\cite{Ji:1996ek,Leader:2013jra,Ji:2020ena,Aidala:2012mv} and to explore the mechanical properties of the hadron~\cite{Polyakov:2002yz,Polyakov:2018zvc,Burkert:2018bqq,Lorce:2018egm,Kumericki:2019ddg,Burkert:2023wzr,Freese:2021qtb,Fujii:2025pkv,Dutrieux:2024bgc,GarciaMartin-Caro:2023toa,Li:2023izn,Lorce:2025ayr,Hu:2024edc,Broniowski:2025ctl}.
GDAs are the $s$-$t$ crossed quantities of the hadronic GPDs,
and both of them can be expressed in terms of the double distributions (DDs)~\cite{Muller:1994ses, Radyushkin:1998bz}. The DDs are often used to construct the GPD model for the extraction  of GPDs from experimental measurements. Thus, the study of GDAs can provide us complementary information on GPDs, and vice versa. The second moments of GDAs give access to the timelike GFFs, from which the spacelike GFFs can be derived via dispersion relations.

As most hadrons are not stable, their GPDs are difficult to probe experimentally. In contrast, the hadronic GDAs can be studied through hadron pair production, which enables measurements for both stable and unstable hadrons. In the reactions $\gamma^{\ast} \gamma  \to  h \bar{h}$ and $\gamma^{\ast} \to h \bar{h} \gamma$, the momentum square of the virtual photon is chosen large enough to satisfy the QCD collinear factorization criteria. The amplitudes of these two-photon reactions can then be expressed in terms of the chiral-even hadronic GDAs, which describe the amplitudes for $q \bar{q} \to h \bar{h}$~\cite{Diehl:1998dk,Diehl:2000uv,Kivel:1999sd,Kumano:2017lhr,Lorce:2022tiq,Lorce:2022cze,Song:2025zwl,Lu:2006ut, Pire:2023kng, Pire:2023ztb,Han:2025mvq}.
 In addition, the authors of Ref.~\cite{Bhattacharya:2025awq} recently proposed that chiral-odd hadronic GDAs can be extracted from the production of  two meson pairs in $e^+e^-$ annihilation. In 2016, the Belle Collaboration reported the first measurement of the cross section for $\gamma^{\ast} \gamma \to \pi^0 \pi^0$~\cite{Belle:2015oin}, from which the quark GDAs and GFFs of the pion were extracted~\cite{Kumano:2017lhr}. With data taking recently underway at Belle II at much higher luminosity, measurements of $\gamma^{\ast} \gamma \to \pi^0 \pi^0$ with unprecedented precision are expected in the near future.

For the production of a baryon-antibaryon pair, the amplitude of $\gamma^{\ast} \to B \bar{B} \gamma$ is expressed in terms of the baryon-antibaryon GDAs, where the virtual photon can come from $e^+e^-$ annihilation, namely, $e^+e^- \to \gamma^* \to B \bar{B} \gamma$, and it is accessible at BESIII  and the  proposed  Super Tau-Charm Facility (STCF)~\cite{Achasov:2023gey}. However, the initial state radiation process also contributes to $e^+e^- \to B \bar{B} \gamma$, and it is described by the timelike electromagnetic  form factors (EM FFs) of the baryon~\cite{Han:2025mvq}.  Since extensive  measurements of the timelike EM FFs of the baryon octet are already available~\cite{CMD-3:2018kql, BESIII:2021tbq,BaBar:2013ves,BESIII:2021rqk,BESIII:2019hdp,BESIII:2022rrg,BESIII:2019tgo,BESIII:2017hyw,BESIII:2019nep,BESIII:2023ioy,BESIII:2020uqk,BESIII:2023ynq,BESIII:2023ldb,Belle:2022dvb,BESIII:2021rkn,BESIII:2021aer,BESIII:2020ktn}, the baryon-antibaryon GDAs can be extracted from the future measurements of  $e^+e^- \to B \bar{B} \gamma$. In this work, we study the baryon–antibaryon GDAs in the reaction $e^{\pm} \gamma \to e^{\pm} B \bar{B}$, which can be measured at Belle II, the proposed STCF, and electron–ion  colliders~\cite{Guo:2025rhh} through ultraperipheral collisions. We also provide the polarization-dependent cross section for the final baryon or antibaryon, since the baryon spin vectors can be determined from their subsequent decays, such as $\Lambda \to N \pi$ and $\Sigma \to N \pi$, where $N$ denotes a nucleon. The polarization-dependent cross section enables the definition of additional spin observables, from which further information on the GDAs can be extracted.

This paper is organized as follows. In Sec.~\ref{kinematics}, we introduce the center-of-mass frame of the baryon–antibaryon pair in the process $e^{\pm} \gamma \to e^{\pm} B \bar{B}$ and define the twist-2 baryon–antibaryon GDAs as well as the baryon EM FFs.
In Sec.~\ref{sacs}, the hadron tensor of $\gamma^{\ast} \gamma \to B \bar{B}$ is expressed in terms of the timelike Compton FFs, and we then derive the corresponding cross sections, including the polarization of the final-state baryon.
Numerical estimates for the cross sections, based on models of the  baryon EM FFs and GDAs, are presented in Sec.~\ref{nucr}.
Finally, our main results are summarized in Sec.~\ref{summary}.

\section{Generalized distribution amplitudes in $e^{\pm}\gamma \to e^{\pm} B \bar{B} $ }
\label{kinematics}
We consider the production of a baryon-antibaryon pair in the following exclusive process,
\begin{align}
e^{\pm}(k_1)\gamma(q_2)\to e^{\pm}(k_2)   B(p_1, S_1) \bar{B}(p_2, S_2),  
\label{eqn:baryon}
\end{align}
where the momenta are indicated in parentheses. We denote the spin vectors of the baryon and antibaryon as
 $S_1$ and $S_2$, respectively, and these spin vectors can be determined from the subsequent decays of hyperons. To describe this process, it is convenient to define the following variables,
\begin{align}
s=(k_1+q_2)^2, \qquad (k_1-k_2)^2=(q_1)^2=-Q^2, \qquad  \hat{s}=P^2=(p_1+p_2)^2, \qquad (p_1)^2=(p_2)^2=m^2,  
\label{eqn:kinva}
\end{align}
where $m$ is the baryon mass.  We choose two lightcone vectors $n$ and $\tilde{n}$, with $n\cdot \tilde n = 1$, which are built  from the photon momenta $q_1$ and $q_2$, as:
\begin{align}
n=\frac{\sqrt{2} Q}{Q^2+\hat{s}} q_2, \qquad \tilde{n}=\frac{\sqrt{2}}{Q} \left(q_1+\frac{Q^2}{Q^2+\hat{s}}q_2\right).
\label{eqn:lcv}
\end{align}
Then, the lightcone components  $v^+=v \cdot n$ and  $v^-=v \cdot \tilde{n}$ are defined for a Lorentz vector $v$.

\begin{figure}[htp]
\centering
\includegraphics[width=0.6\textwidth]{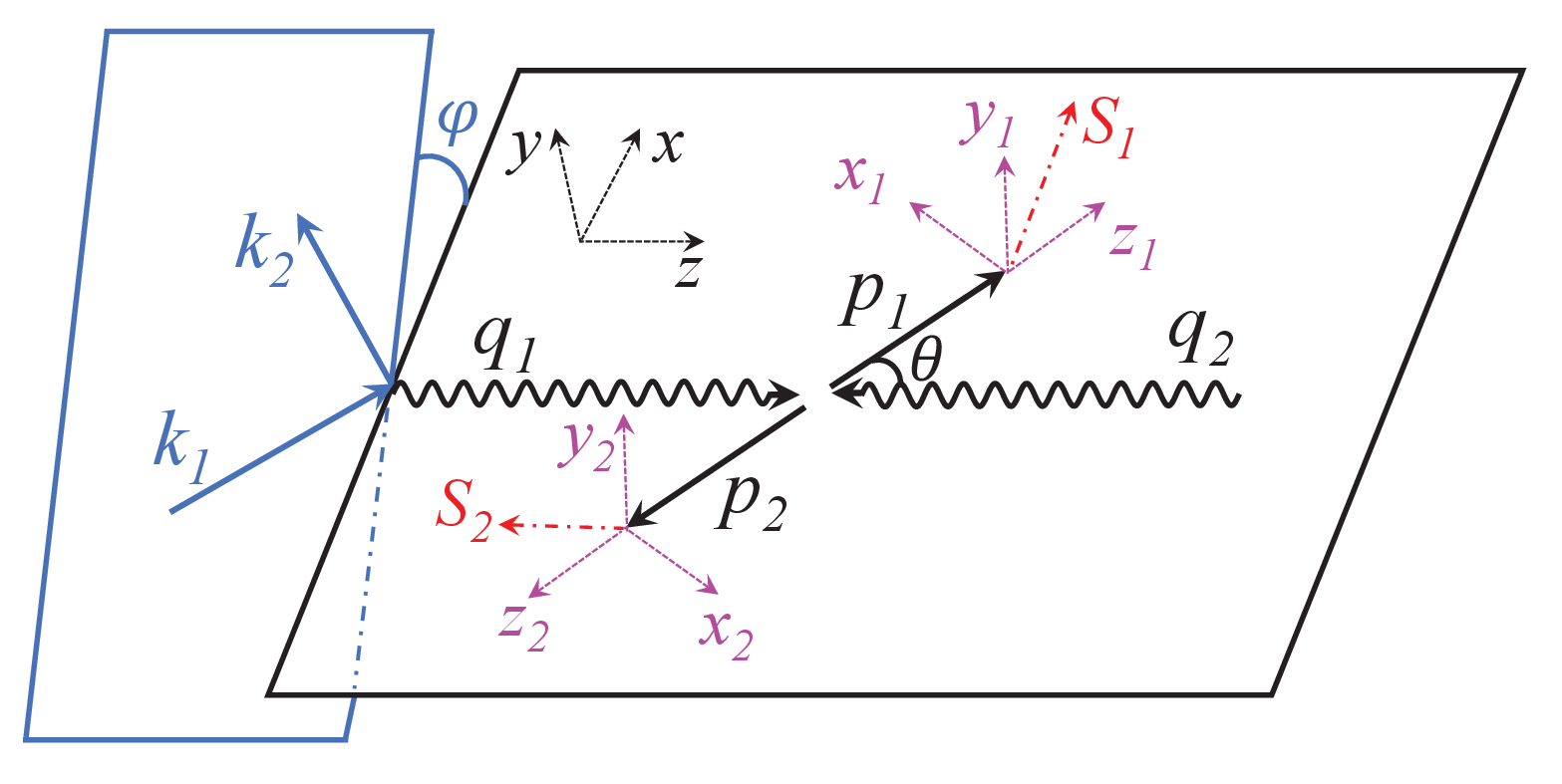}
\caption{
The center-of-mass frame of the baryon-antibaryon pair for the process $e^{\pm}\gamma \to e^{\pm} B \bar{B}$.}
\label{fig:cmf}
\end{figure}

The final cross sections will be presented in the center-of-mass frame of the baryon-antibaryon pair, as described by Fig.~\ref{fig:cmf}. We choose a coordinate system with the $z$ axis  along the momentum $q_1$, and the baryon momenta $p_1$ and $p_2$ lie in the x-z plane,  
The polar angle of $p_1$  is denoted as $\theta$, and $\varphi$ is the azimuthal angle  between the lepton plane and the hadron plane. These angles 
can be expressed in terms of Lorentz invariants,
\begin{equation}
\begin{aligned}
\cos{\theta}=&\frac{q_1\cdot(p_2-p_1)}{\beta_0\,(q_1\cdot q_2)}, \nonumber \\
\sin{\varphi}=&\frac{  4  \epsilon_{\alpha \beta \gamma \delta}  p_1^{\alpha} p_2^{\beta} k_1^{\gamma} q_1^{\delta}  }{\beta_0 \sin{\theta} \sqrt{s Q^2 \hat{s}(s-\hat{s}-Q^2) }},
\label{eqn:pola}
\end{aligned}
\end{equation}
where the convention $\epsilon^{0123}=1$ is used,  and $\beta_0$ is the baryon velocity,
\begin{align}
\beta_0=\sqrt{1-\frac{4 m^2}{\hat{s}}}.
\label{eqn:kinva1}
\end{align}
Note that the spin vectors $S_1$ and $S_2$ are defined within their own rest frames, namely $x_1y_1z_1$ and $x_2y_2z_2$ in Fig.~\ref{fig:cmf}. These vectors are given by 
\begin{equation}
\begin{aligned}
S_1^{\mu}=&(0, S_1^x, S_1^y, S_1^z), \nonumber \\
S_2^{\mu}=&(0, S_2^x, S_2^y, S_2^z).
\label{eqn:spvs}
\end{aligned}
\end{equation}

\begin{figure}[htp]
\centering
\includegraphics[width=0.3\textwidth]{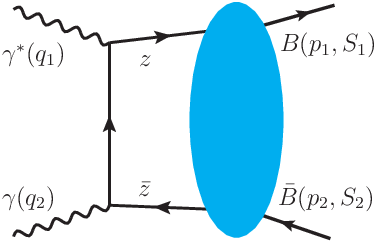}
\caption{
 The Feynman diagram for the QCD subprocess (a)   $ \gamma^{\ast} \gamma \to  B \bar{B}$; the crossed diagram is obtained by interchanging the photon-quark vertices. }
\label{fig:gda}
\end{figure}

In $e^{\pm}\gamma \to e^{\pm} B \bar{B} $, the  $C$-even baryon-antibaryon pairs can be produced in  subprocess (a) where the real photon fuses with the virtual photon emitted by the lepton.
The  amplitude of this subprocess can be expressed in terms of the baryon GDAs using the QCD collinear factorization valid in the generalized Bjorken regime ($Q^2 \gg \hat{s}, \Lambda_{\text{QCD}}^2$) as illustrated by Fig.~\ref{fig:gda}. 
The baryon GDAs are QCD nonperturbative quantities that describe the amplitude for the process $q \bar{q} \to B \bar{B}$.
At leading-twist level, the amplitudes will be investigated using the twist-2 chiral-even GDAs for a baryon-antibaryon pair, which are defined as~\cite{Diehl:2002yh}
\begin{equation}
\begin{aligned}
P^+ \int \frac{dx^-}{2\pi} \,e^{iz P^+  x^-} \langle \bar{B}(p_2) B(p_1)  | \,\bar{q}(-x^-) \gamma^+ q(0)\, | 0 \rangle&
=
\Phi_V^q(z,\zeta_0, \hat{s}) \bar{u}(p_1) \gamma^+ v(p_2)+  \Phi_S^q(z,\zeta_0, \hat{s}) \frac{P^+}{2m} \bar{u}(p_1) v(p_2),\\
P^+ \int \frac{dx^-}{2\pi}\,e^{iz P^+  x^-} \langle \bar{B}(p_2) B(p_1)  | \,\bar{q}(-x^-) \gamma^+ \gamma_5 q(0)\, | 0 \rangle&
=\Phi_A^q(z,\zeta_0, \hat{s}) \bar{u}(p_1) \gamma^+ \gamma_5 v(p_2)+  \Phi_P^q(z,\zeta_0, \hat{s}) \frac{P^+}{2m} \bar{u}(p_1) \gamma_5 v(p_2),
\label{eqn:gdame}
\end{aligned}
\end{equation}
where the lightcone gauge is used, and we omit to write the  factorization scale dependence of the GDAs. In Eq.~\eqref{eqn:gdame}, $z$ is the momentum fraction carried by the quark, and  the skewness parameter $\zeta_0$  is given by
\begin{align}
\zeta_0= \frac{\Delta \cdot n}{P \cdot n}
\label{eqn:skewness}
\end{align}
with $\Delta=p_2-p_1$. This parameter is related to the baryon polar angle by $\zeta_0 = -\beta_0 \cos \theta$.

\begin{figure}[htp]
\centering
\includegraphics[width=0.3\textwidth]{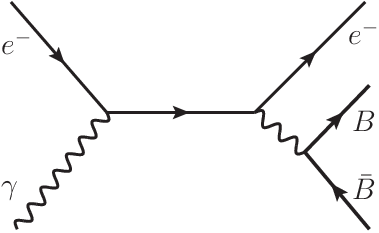}
\caption{
 The Feynman diagram for the bremsstrahlung subprocess (b)  where  the photon-baryon-antibaryon coupling is described by the timelike baryon EM FFs, the crossed diagram is obtained by interchanging the photon-lepton vertices.}
\label{fig:codd}
\end{figure}
The  $C$-odd baryon-antibaryon pairs may originate from a virtual photon emitted by the lepton, which is denoted as subprocess (b) in Fig.~\ref{fig:codd}.
The amplitude of this subprocess can be written as   the timelike baryon EM FFs, 
\begin{align}
 \langle \bar{B}(p_2) B(p_1)  | \,\bar{q}(0) \gamma^{\mu} q(0)\, | 0 \rangle & = F_V^q(\hat{s}) \bar{u}(p_1) \gamma^{\mu} v(p_2) +F_S^q(\hat{s}) \frac{\Delta^{\mu}}{2m} \bar{u}(p_1)  v(p_2), 
\label{eqn:emff}
\end{align}
and these FFs are also related to  the first moments of baryon  GDAs~\cite{Diehl:2002yh}.
In the literature, the electric and magnetic FFs 
$G_E$ and $G_M$ are commonly used, and they can be obtained from a linear combination of $F_V^q(\hat{s})$ and $F_S^q(\hat{s})$,
\begin{equation}
\begin{aligned}
G_E(\hat{s})& =\sum_q e_q \left[ F_V^q(\hat{s}) + (\tau-1)F_S^q(\hat{s}) \right], \\
G_M(\hat{s})&=\sum_q e_q F_V^q(\hat{s}),
\label{eqn:emff12}
\end{aligned}
\end{equation}
with $\tau=\hat{s}/(4m^2)$.

\section{Scattering amplitudes and cross sections }
\label{sacs}
\subsection{Hadron tensor }
\label{crs}

In subprocess (a),  $C$-even baryon-antibaryon pairs are produced, and the corresponding  helicity amplitude  can be written as
\begin{align}
A_{\lambda_1 \lambda_2}=\epsilon_1^{\mu}(\lambda_1) T_{\mu \nu}  \epsilon_2^{\nu}(\lambda_2),
\label{eqn:amp12}
\end{align}
where $\lambda_1$ and $\lambda_2 $ denote the helicities of  the virtual and real photons, respectively. The hadron tensor $T_{\mu \nu}$ is defined by
the  matrix element of  the time-ordered product of two EM currents,
\begin{align}
T_{\mu \nu}=i\int d^4x\, e^{-iq_1\cdot x} \langle \bar{B}(p_2) B(p_1)  | \,
T \{ j_{\mu}^{\text{em}}(x)  j_\nu^{\text{em}} (0) \} \, | 0 \rangle.
\label{eqn:amp0}
\end{align}
At leading twist, the hadron tensor is expressed in terms of four timelike Compton FFs $\mathcal{F}_{i}$,
\begin{align}
T^{\mu \nu}=\frac{-\sqrt{2 } }{ Q} \left \{ g_{T }^{ \mu \nu }\left [ \zeta_0 \mathcal{F}_{V}  \bar{u}  \! \left ( p_{1}  \right )\! \gamma ^{+}  \! v\!\left ( p_{2}  \right )  + \mathcal{F}_{S}  \frac{P^{+} }{ 2m } \bar{u}\!\left ( p_{1}  \right )\! v\!\left ( p_{2}  \right )     \right ] -i\epsilon  _{T}^{\mu \nu} \left [ \mathcal{F}_{A}  \bar{u} \! \left ( p_{1}  \right )\! \gamma ^{+}  \!\gamma ^{5}\! v\!\left ( p_{2}  \right ) +\mathcal{F}_{P}  \frac{P^{+} }{2m } \bar{u}\!\left ( p_{1}  \right )\! \gamma ^{5} \! v\!\left ( p_{2}  \right )   \right ]   \right \},
\label{eqn:tmunu}
\end{align}
and these FFs  are given at leading order in $\alpha_s$ by
\begin{equation}
\begin{aligned}
     \zeta_0 \mathcal{F}_{V} =&\sum_{q} \frac{e_{q}^{2}}{2}\int_{0}^{1} dz\frac{2z-1}{ z\bar{z}  }\Phi _{V}^{q }\!\left ( z,\zeta_0 ,\hat{s}  \right ),  \\
     \mathcal{F}_{S} =&\sum_{q} \frac{e_{q}^{2}}{2}\int_{0}^{1} dz\frac{2z-1}{ z\bar{z}  }\Phi _{S}^{q }\!\left ( z,\zeta_0 ,\hat{s}  \right ),  \\
     \mathcal{F}_{i} =&\sum_{q} \frac{e_{q}^{2}}{2}\int_{0}^{1} dz\frac{1}{ z \bar{z}   }\Phi _{i}^{q }\!\left ( z,\zeta_0,\hat{s}  \right ) \quad \text{for} \quad i=A,\, P,
\label{cffs}
\end{aligned}
\end{equation}
where we introduce the notation $\bar{z} \equiv 1-z$. In Eq.~\eqref{eqn:tmunu}, $ g_{T }^{ \mu \nu }$ and $\epsilon  _{T}^{\mu \nu}$ are transverse tensors,
\begin{equation}
\begin{aligned}
g_{T}^{\alpha \beta} =&g^{\alpha \beta}-
n^{\alpha}\tilde{n}^{\beta}-n^{\beta}\tilde{n}^{\alpha}, \nonumber \\
\epsilon_{T}^{\mu \nu} =& \epsilon^{\mu \nu \alpha \beta} \tilde{n}_{\alpha} n_{\beta},
\label{eqn:gtmu}
\end{aligned}
\end{equation}
The hadron tensor $T_{\mu\nu}$ written above represents the leading-twist contribution at zeroth order in $\alpha_s$. When contracted with the polarization vectors of transversely polarized photons, it leads to the leading-twist amplitudes without a helicity flip. However, the longitudinal virtual photon will contribute to the amplitude if one includes the higher-twist corrections~\cite{Lorce:2022tiq}. At higher order of $\alpha_s$, the gluon GDAs also contribute and the transversity gluon GDAs introduce the third transverse tensor~\cite{Belitsky:2000jk}  $\tau_T^{\mu\nu;\rho\sigma} = (g_T^{\mu\rho}g_T^{\nu\sigma}+ g_T^{\mu\sigma}g_T^{\nu\rho}-g_T^{\mu\nu}g_T^{\rho\sigma})/2$, corresponding to amplitudes with a two-unit photon helicity flip.

Equation \eqref{eqn:tmunu} corresponds to the timelike counterpart of the hadron tensor in Deeply Virtual Compton Scattering (DVCS), in which the spacelike Compton FFs are expressed in terms of GPDs. It has been shown that the extraction of GPDs from DVCS data is an inconclusive endeavor, since so-called “shadow GPDs” yield vanishing contributions to the spacelike Compton FFs, a feature commonly referred to as the deconvolution problem~\cite{Bertone:2021yyz}. In contrast, GDAs are expanded in Gegenbauer ($C^{(i)}_n(2z-1)$) and Legendre ($P_l(\zeta_0)$) polynomials according to their evolution equations~\cite{Diehl:2000uv,Polyakov:1998ze}. Deconvoluting the z-dependence of GDAs therefore does not suffer from the same ambiguities as deconvoluting the x-dependence of GPDs. Moreover, the angular distribution of the baryon–antibaryon pair in the final state provides direct access to the $\zeta_0$-dependence and, albeit indirectly, to the $z$-dependence of the GDAs.

\subsection{Unpolarized cross section}
We sum only over the two polarization states of the antibaryon, and the resulting cross section consists of two components,
\begin{equation}
\frac{d\sigma(S_1)  }{d\hat{s}dQ^{2} d\left ( \cos \theta  \right ) d\varphi    }= \frac{1}{2} \frac{d \bar{\sigma}  }{d\hat{s}dQ^{2} d\left ( \cos \theta  \right ) d\varphi    } +\frac{d \hat{\sigma}(S_1) }{d\hat{s}dQ^{2} d\left ( \cos \theta  \right ) d\varphi    }.
\label{eqn:crospin}
\end{equation}
The first term represents the unpolarized cross section, which remains after summing over the polarization states of the baryon. The second term depends on the baryon spin vector $S_1$ and corresponds to the single-spin correlation.

We introduce a few dimensionless parameters to describe the cross sections for $e^{\pm}\gamma \to e^{\pm} B \bar{B}$,
\begin{align}
\epsilon=\frac{1-y}{1-y+\frac{y^2}{2}}, \qquad y=\frac{q_1\cdot q_2}{k_1\cdot q_2}, \qquad x= \frac{Q^2}{Q^2+\hat{s}},\qquad \lambda= \frac{1}{(\beta_0)^2}.
\label{eqn:polar}
\end{align}
Given that the cross sections become very cumbersome, particularly when baryon polarizations are included, it is convenient to define some coefficients expressed in terms of $\epsilon$ and $x$,
\begin{equation}
\begin{aligned}
\omega_1&=1-2x\left ( 1-x \right )  \left ( 1-\epsilon  \right ),  &\omega_5&=\sqrt{\epsilon \left ( 1+\epsilon  \right ) }\sqrt{2x\left ( 1-x \right ) }\left ( 2x-1 \right )  ,\\
\omega_2&=1-2x\left ( 1-x \right )  \left ( 1+\epsilon  \right ), &\omega_6&=\sqrt{\epsilon \left ( 1+\epsilon  \right ) }\sqrt{2x\left ( 1-x \right ) }, \\
\omega_3&=2\epsilon x\left ( 1-x \right ),     &\omega_7&=1-\left ( 1-x \right )  \left ( 1-\epsilon  \right ), \\
\omega_4&=1-2x\left ( 1-x \right ),    &\omega_8&=1-\left ( 1-x \right )  \left ( 1+\epsilon  \right ).
\label{eqn:coisr01}
\end{aligned}
\end{equation}

We first calculate the unpolarized cross sections, and   the contribution of subprocess (a) to the differential cross section  is obtained using the hadron tensor of Eq.~\eqref{eqn:tmunu},
\begin{equation}
\begin{aligned}
\frac{d\bar{\sigma} _{\mathrm{G} } }{d\hat{s}dQ^{2} d( \cos \theta) d\varphi    } = & \frac{\alpha _{\mathrm{em} }^{3} \beta _{0}}{8\pi s^{2} Q^{2}   }\frac{1}{1-\epsilon }  \Big \{   \left  | \mathcal{F} _{A}  \right |^{2} -\left | \mathcal{F} _{S}  \right | ^{2}+2\mathrm{Re} \left ( \mathcal{F} _{A}\mathcal{F} _{P}^{\ast } \right)+\frac{\hat{s}  }{4m^{2}}  ( \left | \mathcal{F} _{P}  \right |^{2} +\left | \mathcal{F} _{S}  \right |^{2}   ) \\  & + (\beta _{0})^2 \cos^2\theta \left[  \left | \mathcal{F} _{V}  \right |^{2} +  2\mathrm{Re} \left ( \mathcal{F} _{S}\mathcal{F} _{V}^{\ast } \right )- \left | \mathcal{F} _{A}  \right | ^{2} \right]   -(\beta _{0})^4 \cos^4\theta \left | \mathcal{F} _{V}  \right |^{2}   \Big \},
\label{crogda}
\end{aligned}
\end{equation}
where the four twist-2 GDAs appear in the timelike Compton FFs, and this GDA contribution is independent on the azimuthal angle $\varphi$. It is straightforward to calculate the contribution of the bremsstrahlung subprocess (b), which  are described by  the timelike baryon EM FFs of Eq.~\eqref{eqn:emff12},
\begin{equation}
\begin{aligned}
\frac{d\bar{\sigma }_{\mathrm{B} } }{d\hat{s}dQ^{2} d( \cos \theta) d\varphi    }=&\frac{\alpha _{\mathrm{em} }^{3}\beta_{0}^{3}   }{4\pi s^{2} }\frac{1}{\epsilon \hat{s} }  \Big \{ \omega_1 (2\lambda-1)  \left | G_{M}  \right |^{2} + \left [  \omega_2\left | G_{M}\right |^{2}+2\omega_3 (\lambda-1) ( \left | G_{E} \right |^{2} -\left |G_{M}\right |^{2} )\right] \cos ^{2} \theta  \\
&+\left[  \omega_3\left | G_{M}\right |^{2}+\omega_4 (\lambda-1) ( \left | G_{E} \right |^{2} -\left |G_{M}\right |^{2} )   \right] \sin ^{2}\theta +
\omega_5 \left[   (\lambda-1)  \left | G_{E} \right |^{2} -\lambda \left |G_{M}\right |^{2}   \right]  \\
&\times \sin \left ( 2\theta  \right )\cos \varphi -\omega_3 \left[   (\lambda-1)  \left | G_{E} \right |^{2} -\lambda \left |G_{M}\right |^{2}   \right] \sin ^{2}\theta \cos \left ( 2\varphi  \right )   \Big \},
\label{eqn:isr-tes}
\end{aligned}
\end{equation}
and the dependence on the azimuthal angle appears in this contribution. Then, the interference contribution of the two amplitudes of subprocesses (a) and (b) is written as
\begin{equation}
\begin{aligned}
\frac{d\bar{\sigma}_{\mathrm{I} } }{d\hat{s}dQ^{2} d( \cos \theta) d\varphi  } =&e_l \frac{\alpha  _{\mathrm{em} }^{3} \beta _{0}  }{8\pi s^{2} }\frac{\sqrt{2}\beta _{0}  }{Q \sqrt{\hat{s}  \epsilon (1-\epsilon )} } \Big \{2\omega_6  \left[ \mathrm{Re} \left (  G_{M}^{\ast } \mathcal{F}_{V} \right ) +\mathrm{Re}(G_{E}^{\ast }\mathcal{F}_{S} ) \right]\cos \theta -2(\beta_0)^2 \omega_6 \left[\lambda \mathrm{Re}(G_{M}^{\ast }\mathcal{F}_{V} ) \right. \\
&\left.-(\lambda-1) \mathrm{Re}\left( G_{E}^{\ast }\mathcal{F}_{V}\right )  \right]\cos ^{3}\theta
+2\left [ \omega_7 \mathrm{Re} ( G_{M}^{\ast }\mathcal{F} _{A}    ) + \omega_8 \mathrm{Re}(G_{E}^{\ast }\mathcal{F}_{S} ) \right] \sin \theta \cos \varphi\\
&-(\beta_0)^2 \omega_8\left[ \lambda \mathrm{Re}(G_{M}^{\ast }\mathcal{F}_{V} ) -(\lambda-1) \mathrm{Re}\left( G_{E}^{\ast }\mathcal{F}_{V}\right ) \right]
\sin \left ( 2\theta  \right ) \cos \theta \cos \varphi  \Big \},
\label{intcross}
\end{aligned}
\end{equation}
where $e_l$ is the charge of the lepton $e^{\pm}$. After integration over azimuthal angle, the differential cross sections are expressed as
\begin{equation}
\begin{aligned}
\frac{d\bar{\sigma} _{\mathrm{G} } }{d\hat{s}dQ^{2} d( \cos \theta)    } = & \frac{\alpha _{\mathrm{em} }^{3} \beta _{0}}{4 s^{2} Q^{2}   }\frac{1}{1-\epsilon }  \Big \{   \left  | \mathcal{F} _{A}  \right |^{2} -\left | \mathcal{F} _{S}  \right | ^{2}+2\mathrm{Re} \left ( \mathcal{F} _{A}\mathcal{F} _{P}^{\ast } \right)+\frac{\hat{s}  }{4m^{2}} \left ( \left | \mathcal{F} _{P}  \right |^{2} +\left | \mathcal{F} _{S}  \right |^{2}   \right ) \\  & + (\beta _{0})^2 \cos^2\theta \left[  \left | \mathcal{F} _{V}  \right |^{2} +  2\mathrm{Re} \left ( \mathcal{F} _{S}\mathcal{F} _{V}^{\ast } \right )- \left | \mathcal{F} _{A}  \right | ^{2} \right]   -(\beta _{0})^4 \cos^4\theta \left | \mathcal{F} _{V}  \right |^{2}   \Big \},\\
\frac{d\bar{\sigma }_{\mathrm{B} } }{d\hat{s}dQ^{2} d( \cos \theta)    }=&\frac{\alpha _{\mathrm{em} }^{3}\beta_{0}^{3}   }{2s^{2} }\frac{1}{\epsilon \hat{s} }  \Big \{ \omega_1 (2\lambda-1)  \left | G_{M}  \right |^{2} + \left [  \omega_2\left | G_{M}\right |^{2}+2\omega_3 (\lambda-1) ( \left | G_{E} \right |^{2} -\left |G_{M}\right |^{2} )\right]\cos ^{2} \theta  \\&
  +\left[  \omega_3\left | G_{M}\right |^{2}+\omega_4 (\lambda-1) ( \left | G_{E} \right |^{2} -\left |G_{M}\right |^{2} )   \right] \sin ^{2}\theta \Big \},\\
\frac{d\bar{\sigma}_{\mathrm{I} } }{d\hat{s}dQ^{2} d( \cos \theta)  } =&e_l \frac{\alpha  _{\mathrm{em} }^{3} \beta _{0}  }{4s^{2} }\frac{\sqrt{2}\beta _{0}  }{Q \sqrt{\hat{s}  \epsilon (1-\epsilon )} } \Big \{2\omega_6  \left[ \mathrm{Re} \left ( G_{M}^{\ast }\mathcal{F}_{V}  \right ) +\mathrm{Re}( G_{E}^{\ast } \mathcal{F}_{S}) \right]\cos \theta \\&-2(\beta_0)^2 \omega_6 \left[\lambda \mathrm{Re}(G_{M}^{\ast } \mathcal{F}_{V} ) \right. \left.-(\lambda-1) \mathrm{Re}\left( G_{E}^{\ast } \mathcal{F}_{V}\right )  \right]\cos ^{3}\theta  \Big \}.
\label{crointph}
\end{aligned}
\end{equation}

In the cross sections, the physical quantities we are interested in are the baryon GDAs, which only exist in Eqs.~\eqref{crogda} and \eqref{intcross}.  If we take the difference of the cross sections in the processes  $e^{-}\gamma \to e^{-} B \bar{B}$ and $e^{+}\gamma \to e^{+} B \bar{B}$, only the interference term will remain. 
Thus, we can employ the lepton charge asymmetry to extract the GDAs from the interference term, in close analogy to the deeply virtual Compton scattering. In addition, one can make use of the exchange of $(\theta, \varphi) \to (\pi-\theta, \pi+\varphi)$ in the cross sections. The charge conjugation of the baryon–antibaryon pairs differs between subprocesses (a) and (b). Consequently, Eqs.~\eqref{crogda} and \eqref{eqn:isr-tes} remain unchanged under this exchange, whereas Eq.~\eqref{intcross} changes sign.
We can define a asymmetry as
\begin{equation}
    \bar{A}(\theta,\varphi) =\frac{\frac{d\bar{\sigma} (\theta,\varphi)}{d\cos \theta \, d \varphi}- \frac{d\bar{\sigma} (\pi -\theta,\varphi+\pi)}{d\cos \theta \, d \varphi}}{\frac{d\bar{\sigma} (\theta,\varphi)}{d\cos \theta \, d \varphi}+ \frac{d\bar{\sigma} (\pi -\theta,\varphi+\pi)}{d\cos \theta \, d \varphi}},
\end{equation}
and only the interference contribution survives in the numerator. Furthermore, a forward-backward asymmetry may be defined by integrating the azimuthal angle $\varphi$ from $\pi/2$ to $3\pi/2$ and subtracting the integral over the complementary region, 
\begin{equation}
    \bar{A}_{FB} (\theta) =\frac{\int_{\pi/2}^{3\pi/2} d\varphi\frac{d\bar{\sigma} (\theta,\varphi)}{d\cos \theta ~ d \varphi}- \int_{3\pi/2}^{2\pi} d \varphi\frac{d\bar{\sigma} (\pi -\theta,\varphi)}{d\cos \theta ~ d \varphi}- \int_{0}^{\pi/2} d \varphi\frac{d\bar{\sigma} (\pi -\theta,\varphi)}{d\cos \theta ~ d \varphi}}{\int_{\pi/2}^{3\pi/2} d\varphi\frac{d\bar{\sigma} (\theta,\varphi)}{d\cos \theta ~ d \varphi}+ \int_{3\pi/2}^{2\pi} d \varphi\frac{d\bar{\sigma} (\pi -\theta,\varphi)}{d\cos \theta ~ d \varphi}+ \int_{0}^{\pi/2} d \varphi\frac{d\bar{\sigma} (\pi -\theta,\varphi)}{d\cos \theta ~ d \varphi}}\,.
\end{equation}
The numerator of $\bar{A}_{FB}$ is denoted as $\int_{\Delta \varphi} d\varphi \frac{d\Delta \bar{\sigma} (\theta,\varphi)}{d\cos \theta ~ d \varphi}$, which is expressed as
\begin{equation}
\begin{aligned}
 \frac{d\bar{\sigma}_{\mathrm{FB} } }{d\hat{s}dQ^{2} d\left ( \cos \theta  \right )    }=&\int_{\Delta\varphi} d\varphi\frac{d\Delta\bar{\sigma}(\theta,\varphi)}{d\cos \theta ~ d \varphi}   \\
 = &e_l \frac{\alpha  _{\mathrm{em} }^{3} \beta _{0}  }{2\pi s^{2} }\frac{\sqrt{2}\beta _{0}  }{Q \sqrt{\hat{s}  \epsilon (1-\epsilon )} } \Big \{ \pi\omega_6  \left[ \mathrm{Re} \left ( G_{M}^{\ast }\mathcal{F}_{V}  \right ) +\mathrm{Re}(G_{E}^{\ast } \mathcal{F}_{S} ) \right]\cos \theta -\pi(\beta_0)^2 \omega_6 \left[\lambda \mathrm{Re}(G_{M}^{\ast} \mathcal{F}_{V} ) \right. \\
&\left.-(\lambda-1) \mathrm{Re}\left( G_{E}^{\ast }\mathcal{F}_{V}\right )  \right]\cos ^{3}\theta
-2\left [ \omega_7 \mathrm{Re} ( G_{M}^{\ast }\mathcal{F} _{A}    ) + \omega_8 \mathrm{Re}( G_{E}^{\ast }\mathcal{F}_{S}) \right] \sin \theta \\
&+(\beta_0)^2 \omega_8\left[ \lambda \mathrm{Re}( G_{M}^{\ast }\mathcal{F}_{V}) -(\lambda-1) \mathrm{Re}\left( G_{E}^{\ast }\mathcal{F}_{V}\right ) \right]
\sin \left ( 2\theta  \right ) \cos \theta  \Big \}.
\label{intasym}
\end{aligned}
\end{equation}
This yields a new observable for accessing the baryon GDAs from the interference term.

\subsection{Single-spin correlation}
We now present the single-spin correlation for $e^{\pm}\gamma \to e^{\pm} B \bar{B}$. The pure QCD contribution is 
written as the timelike Compton FFs,
\begin{equation}
     \frac{d\hat{\sigma}  _{G}(S_{1} ) }{d\hat{s}dQ^{2} d(\cos\theta)  d\varphi  } =\frac{\alpha _{\mathrm{em} }^{3} \beta _{0}^{2}    }{32\pi s^{2} Q^{2} (1-\epsilon ) }\frac{\sqrt{\hat{s} } }{m}\left [ -\beta _{0}  \sin(2\theta)  \,\mathrm{Im}\left ( \mathcal{F}_{V}\mathcal{F} _{S}^{\ast }     \right )+ 2\sin\theta \, \mathrm{Im}\left ( \mathcal{F} _{A} \mathcal{F} _{P}^{\ast }   \right )   \right ]{S}_{1}^{y}.
 \label{gdaspin}    
\end{equation}
In contrast, its counterpart, Eq.~\eqref{crogda}, probes the real parts of the products of two Compton FFs  $\mathcal{F}_{i}\mathcal{F}^{*}_{j}$, while by including baryon polarizations, we gain access to their imaginary parts. This provides complementary information to the unpolarized cross sections. In addition, we  can also sum over the two polarization states of the baryon and obtain the single-spin correlation that is dependent on the spin vector $S_2$ of the antibaryon using the following replacements,
\begin{equation}
\begin{aligned}
(\theta, \varphi) &\to (\pi-\theta, \pi+\varphi), \\
(S_1^x, S_1^y, S_1^z) &\to (-S_2^x, -S_2^y, S_2^z).
\label{eq:reps}
\end{aligned}
\end{equation}
Then,  the second term in Eq.~\eqref{gdaspin} changes sign for the $S_2$ 
case, whereas the sign of the first term remains unchanged.
Thus, one can infer  that the spin vector of the baryon or antibaryon is always perpendicular to the hadron plane, as illustrated in Fig.~\ref{fig:cmf}. 
The first term in Eq.~\eqref{gdaspin} induces a parallel alignment between the spin vectors of the produced baryon and antibaryon, while the second term leads to an antiparallel alignment.

The contribution of the bremsstrahlung subprocess to the differential cross section is expressed as  
\begin{equation}
\begin{aligned}              
  \frac{d\hat{\sigma} _{B}(S_{1} ) }    {d\hat{s}dQ^{2}d(\cos\theta )d\varphi    } =& \frac{\alpha _{\mathrm{em} }^{3}\beta _{0}   }{4\pi s^{2} \epsilon \hat{s} }  \frac{m}{\sqrt{\hat{s} } }  \mathrm{Im}(G_{M} G_{E}^{\ast } ) \Big \{ 2\big[\omega _{3}\sin\theta\sin(2\varphi)  -\omega _{5}\cos\theta \sin\varphi  \big]  {S}_{1}^{x}\\&+\big[\omega _{2}\sin(2\theta)+2\omega _{5}\cos(2\theta) \cos\varphi   -2\omega _{3}\sin(2\theta)  \cos^{2}\!\varphi   \big]{S}_{1}^{y}\  \!\Big \},  
\label{emspin}  
\end{aligned}
\end{equation}
where the coefficients $\omega _{i}$ are defined by Eq.~\eqref{eqn:coisr01}.
Similarly, we can obtain the single-spin correlation for the $S_2$ case using the replacements in Eq.~\eqref{eq:reps}.
The spin vector of the baryon (or antibaryon) lies entirely within the $x$-$y$ plane of its rest frame. 
Finally, we calculate the interference contribution of subprocesses (a) and (b), 
\begin{equation}
\begin{aligned}
\frac{d\hat{\sigma} _{\mathrm{I} }(S_{1} ) }{d\hat{s}dQ^{2} d\left ( \cos \theta  \right ) d\varphi  } =& \frac{e_{l}\alpha  _{\mathrm{em} }^{3} \beta _{0}   }{8\pi s^{2}Q \sqrt{2 \hat{s}  \epsilon (1-\epsilon )}}  \Bigg  \{\bigg [ \beta _{0}\Big(\omega _{8}\mathrm{Im}(G_{M}^{\ast } \mathcal{F}_{S}  ) \sin\varphi + \omega_{8}\frac{4m^{2} }{\hat{s} } \mathrm{Im}(G_{M}^{\ast } \mathcal{F}_{V}  )\cos^{2} \!\theta \sin\varphi\\&+\omega _{7} \frac{4m^{2} }{\hat{s} } \mathrm{Im}(G_{E}^{\ast } \mathcal{F}_{A}  )   \sin^{2} \!\theta \sin\varphi  \Big)+\omega _{7}\mathrm{Im}\big (G_{M}^{\ast } (\mathcal{F}_{P}+\frac{4m^{2} }{\hat{s} } \mathcal{F} _{A}   )\big ) \cos\theta\sin\varphi \bigg] \frac{\sqrt{\hat{s} } }{m}S_{1}^{x} \\&+\bigg[\beta _{0}\Big(\omega_{6}\mathrm{Im}(G_{M}^{\ast } \mathcal{F}_{S}  ) \sin\theta -\omega_{8}\mathrm{Im}\big(G_{M}^{\ast } (\mathcal{F}_{S}+\frac{4m^{2} }{\hat{s} } \mathcal{F} _{V}) \big)  \cos\theta \cos\varphi+\omega_{6} \frac{2m^{2} }{\hat{s} } \\&\times \mathrm{Im}\big( (G_{M}^{\ast }-G_{E}^{\ast } )\mathcal{F}_{V} \big) \sin(2\theta) \cos\theta+\omega_{8}\frac{2m^{2} }{\hat{s} } \mathrm{Im}\big( (G_{M}^{\ast }-G_{E}^{\ast } ) \mathcal{F}_{V}\big)\sin(2\theta) \sin\theta\cos\varphi\Big)\\&-\omega _{7} \mathrm{Im}\big (G_{M}^{\ast } (\mathcal{F}_{P}+\frac{4m^{2} }{\hat{s} } \mathcal{F} _{A}   )\big) \cos\varphi\bigg ]\frac{\sqrt{\hat{s} } }{m} S_{1}^{y}  +\bigg[2\omega _{7} \mathrm{Im}\big(G_{E}^{\ast } (\mathcal{F}_{P}+\frac{4m^{2} }{\hat{s} } \mathcal{F} _{A} )\big) \sin\theta\sin\varphi   \\&-\beta _{0} \big(  \omega _{7} \mathrm{Im}( G_{M}^{\ast }\mathcal{F }_{A})- \omega _{8}  \mathrm{Im}( G_{M}^{\ast }\mathcal{F }_{V} ) \big)  \sin(2\theta )\sin\varphi  \bigg]S_{1}^{z} \Bigg \}.
\label{sscin}
\end{aligned}
\end{equation}
To obtain the counterpart of Eq.~\eqref{sscin} that depends on $S_2$, an additional minus sign must be included following the replacements in Eq.~\eqref{eq:reps}.  The interference term shows that the produced baryon or antibaryon can be polarized along the $z$-axis
of its rest frame, which does not exist in Eqs.~\eqref{gdaspin} and \eqref{emspin}.

\section{Numerical estimates of  the cross sections}
\label{nucr}

In this section, we present  numerical estimates for $e^{\pm} \gamma \to  e^{\pm} p  \bar{p}$ cross sections and asymmetries, using models of the proton EM FFs based on experimental information and a simple but motivated model for proton GDAs derived in our former work~\cite{Han:2025mvq}. This will provide guidance for future experimental measurements that will ultimately lead to more realistic models of GDAs. 

Much experimental information exists on the modulus of the proton EM FFs on the spacelike and timelike regions~\cite{Liu:2015jna,Afanasev:2017gsk,Bytev:2019rdc,Lin:2021xrc,McRae:2023zgu,Galynskii:2024thr,Kuzmin:2024ozz, Bianconi:2015owa,Tomasi-Gustafsson:2020vae,Lomon:2012pn,Qian:2022whn,Yan:2023nlb,Huang:2021xte,  Yang:2024iuc, Cao:2021asd}. As discussed in~\cite{Han:2025mvq}, we  neglect  the unknown phases of the FFs and use a two component description~\cite{Tomasi-Gustafsson:2020vae} that fits experimental data in a satisfactory way. In this work, we do not repeat the parameterization of proton EM FFs, as the details can be found in Refs.~\cite{Han:2025mvq,Tomasi-Gustafsson:2020vae}.

Our model~\cite{Han:2025mvq} for the nucleon-antinucleon GDAs is based on an analogy with the measured $\pi \pi$ GDAs and the decomposition coming from the evolution equations (which are the same for both GDAs). The phases of the $N \bar N$ GDAs are unknown and we neglect their effects, while the moduli are constrained by the sum rules which relate them to the quark momentum content of the hadron (the $R_q$ measuring the ratio of momentum carried by quark $q$ in the hadron, as measured in deep inelastic scattering). This allows us to estimate the proton Compton FFs $ \mathcal{F}_{S}$ and $ \mathcal{F}_{V}$ as:
\begin{equation}
\begin{aligned}
\mathcal{F}_{S} &  =\sum_{q} \frac{e_{q}^{2}}{2}\int_{0}^{1} dz\frac{2z-1}{ z\bar{z}  }\Phi _{N \bar N}^{q }\!\left ( z,\zeta_0 ,\hat{s}  \right ), \\
\zeta_0 \mathcal{F}_{V}  & =\mathcal{F}_{S}\cos \theta ,
\label{eqn:cffpionp}
\end{aligned}
\end{equation}
with
\begin{align}
\Phi^q_{N\bar N }(z, \cos\theta,  \hat{s})  =
 \frac{3 (2 \alpha +3 )}{5B(\alpha+1,\alpha+1)} z^\alpha(1-z)^\alpha (2z-1)
\left [ \widetilde{B}_{10}^q(\hat{s})+\widetilde{B}_{12}^q(\hat{s})P_2(\cos\theta) \right],
\label{eqn:phi-paramet1}
\end{align}
where $\alpha = 1.157$ is obtained from fits to $\pi \pi$ experimental data and is very close to the asymptotic prediction $\alpha =1$.
$B(a, b)$ are Euler's beta functions.
In our simple model, we neglect the imaginary phases for 
the complex functions $\widetilde{B}_{nl}$ and write,
\begin{equation}
\begin{aligned}
\widetilde{B}_{10}^q(\hat{s})& = - \frac{10R_{q}}{9}\left( 1 +\frac{2m^2}{\hat{s}} \right) \, F_q(\hat{s}), \\
\widetilde{B}_{12}^q(\hat{s})& = \frac{10R_{q}}{9} \left( 1 -\frac{4m^2}{\hat{s}} \right) \, F_q (\hat{s}),
\label{eqn:res}
\end{aligned}
\end{equation}
where $R_u=1/3$, $R_d=1/6$ , and $m$ is the proton mass instead of the pion mass. $ F_q (\hat{s}) $ is the monopole FF,
\begin{align}
 F_q (\hat{s}) = \frac{1}{\left[ 1 + (\hat{s}-4 m^2)/\Lambda^2 \right]},
\end{align}
where $\Lambda=1.928$ GeV is the cut-off parameter. We also neglect resonance contributions in the GDAs, which are expected to play an important role in the resonance regions.
The Compton FFs $\mathcal{F}_{A}$ and $\mathcal{F}_{P}$ are related to the axial-vector GDAs $\Phi_{A}^{q}$ and $\Phi_{P}^{q}$, respectively.
Since no studies on the axial-vector GDAs  are currently available, we assume that $\mathcal{F}_{A}$ and $\mathcal{F}_{P}$ are  proportional to the axial charge,
\begin{align}
\mathcal{F}_{A} = \mathcal{F}_{P} = g_A \mathcal{F}_{S},
\label{eqn:sdep-ff1}
\end{align}
with 
\begin{align}
g_A=e_u g_A^u +e_d g_A^d.
\end{align}
In this work, we adopt  $g_A^u=0.832$ and $g_A^d=-0.417$ for the axial charges~\cite{Alexandrou:2024ozj}. 

\begin{figure}[htb]
\centering
\includegraphics[width=0.9\textwidth]{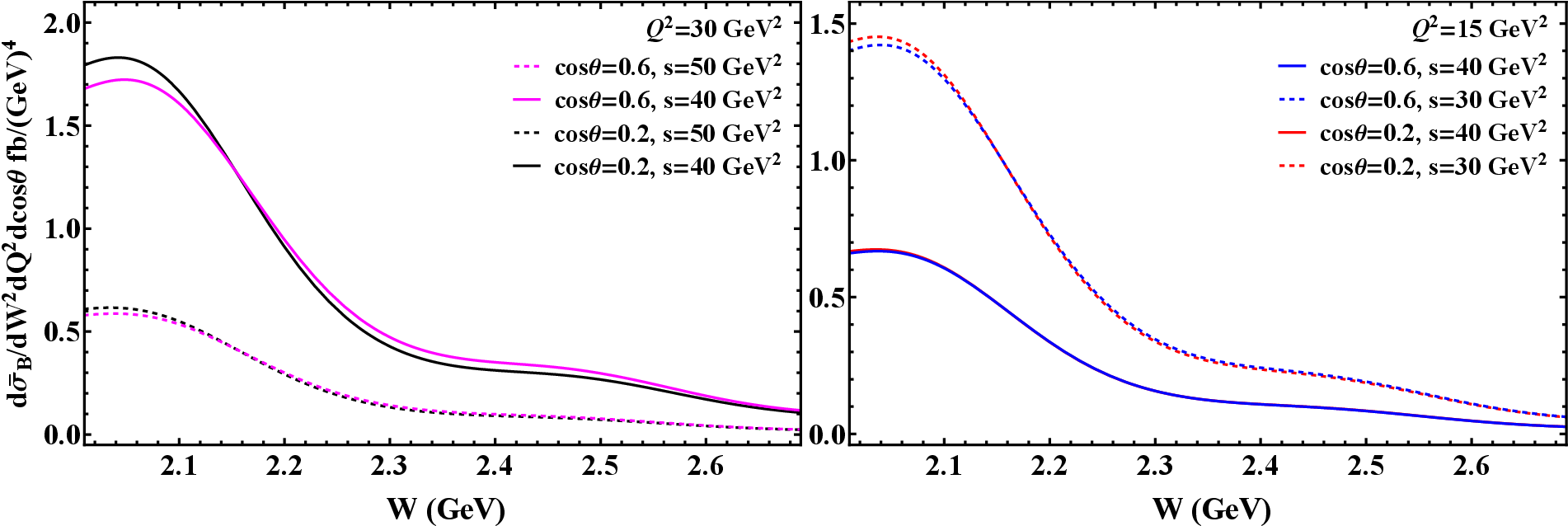}
\caption{ The differential cross section of the bremsstrahlung process in $e^- \gamma  \to  e^-  p \bar{p} $.}
\label{fig:crob}
\end{figure}

In our numerical estimates, the kinematics are chosen according to Belle II for $e^- \gamma  \to  e^-  p \bar{p} $. We use the differential cross sections after integration over azimuthal angle, which are given in Eqs.~\eqref{crointph} and \eqref{intasym}. In Fig.~\ref{fig:crob}, the cross section of the bremsstrahlung process is shown using the model for the proton EM FFs. In the left panel, we set $Q^2 = 30~\text{GeV}^2$ and take $s = 40~\text{GeV}^2$ and $50~\text{GeV}^2$ as typical values, where the black (magenta) curves correspond to $\cos\theta =  0.2$ ($ 0.6$). The invariant mass of the proton–antiproton pair is chosen within the range $2.0~\text{GeV} < W = \sqrt{\hat{s}} < 2.7~\text{GeV}$. The bremsstrahlung cross section shows  a weak dependence on the polar angle $\cos\theta$. In the right panel of Fig.~\ref{fig:crob}, we reduce $Q^2$ from $30~\text{GeV}^2$ to $15~\text{GeV}^2$, where $s = 30~\text{GeV}^2$ and $40~\text{GeV}^2$ are used, and the red (blue) curves correspond to $\cos\theta =  0.2$ ($ 0.6$).
The bremsstrahlung contribution varies significantly when comparing the results at $Q^2 = 15~\text{GeV}^2$ and $Q^2 = 30~\text{GeV}^2$ in Fig.~\ref{fig:crob} for fixed $s = 40~\text{GeV}^2$. This variation originates from the parameter $\epsilon$. For instance, we have $\epsilon = 0.81$ at $W = 2.1~\text{GeV}$, $Q^2 = 15~\text{GeV}^2$, and $s = 40~\text{GeV}^2$, while increasing $Q^2$ to $30~\text{GeV}^2$ reduces $\epsilon$ to $0.27$. Consequently, according to Eq.~\eqref{crointph}, the corresponding cross section becomes about three times larger, in agreement with the behavior observed in Fig.~\ref{fig:crob}. Note that our predictions of  the bremsstrahlung contribution can be considered as accurate estimates, since only 
the modulus of the timelike proton electric and magnetic FFs are used as inputs, which have been well constrained by experiment.

\begin{figure}[htb]
\centering
\includegraphics[width=0.9\textwidth]{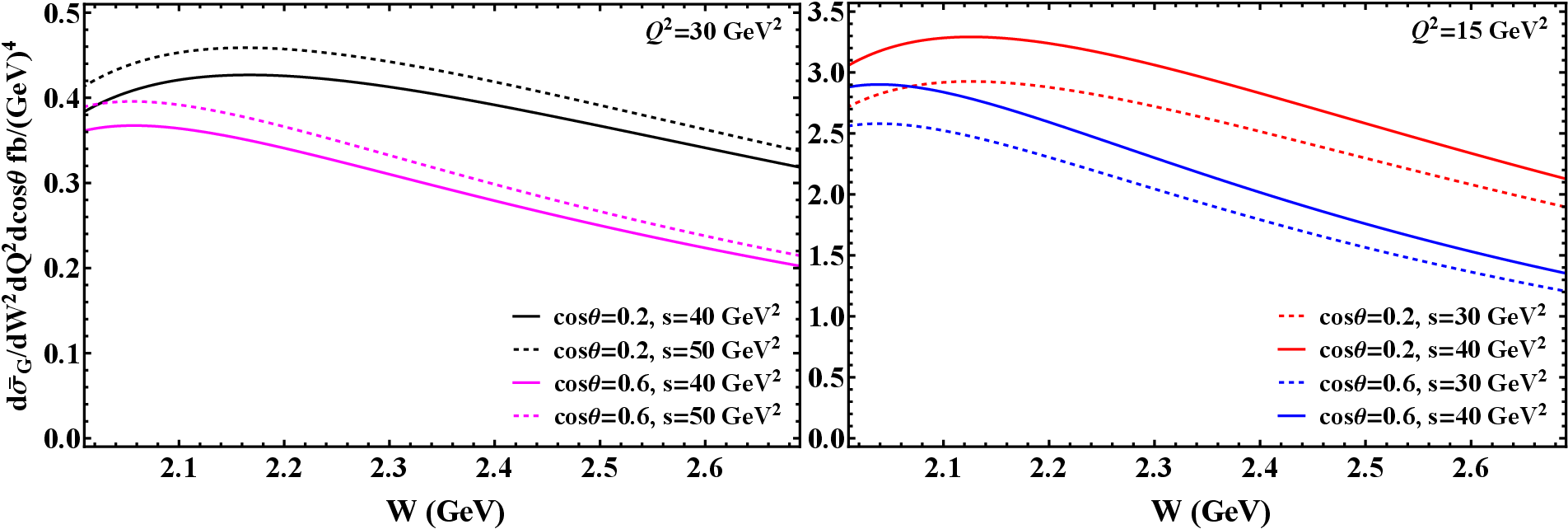}
\caption{ The differential cross section of the pure QCD contribution in $e^- \gamma  \to  e^-  p \bar{p} $.}
\label{fig:crossg}
\end{figure}

The differential cross section of the pure QCD process is depicted  in Fig.~\ref{fig:crossg}, where the model of proton-antiproton GDAs is used. The same conventions are adopted as  Fig.~\ref{fig:crob}, and we choose  $Q^2=30~\text{GeV}^2$ and $15~\text{GeV}^2$ in the right and left panels, respectively.
Compared with the bremsstrahlung contribution shown in Fig.~\ref{fig:crob}, the GDA contribution to the process $e^- \gamma \to e^- p \bar{p}$ is smaller at $Q^2 = 30~\text{GeV}^2$. However, the GDA contribution increases significantly when $Q^2$ decreases from $30~\text{GeV}^2$ to $15~\text{GeV}^2$. This enhancement arises from the prefactor $1/[Q^2(1-\epsilon)]$ in Eq.~\eqref{crointph}. As $Q^2$ changes from $30~\text{GeV}^2$ to $15~\text{GeV}^2$ with fixed $W = 2.1~\text{GeV}$ and $s = 40~\text{GeV}^2$, this factor increases by approximately an order of magnitude, which is consistent with our numerical results for the GDA contribution. Thus, we can find a kinematical region where the GDA contribution may be larger than the bremsstrahlung contribution, which should be helpful for the extraction of the baryon GDAs.

\begin{figure}[htb]
\centering
\includegraphics[width=0.90\textwidth]{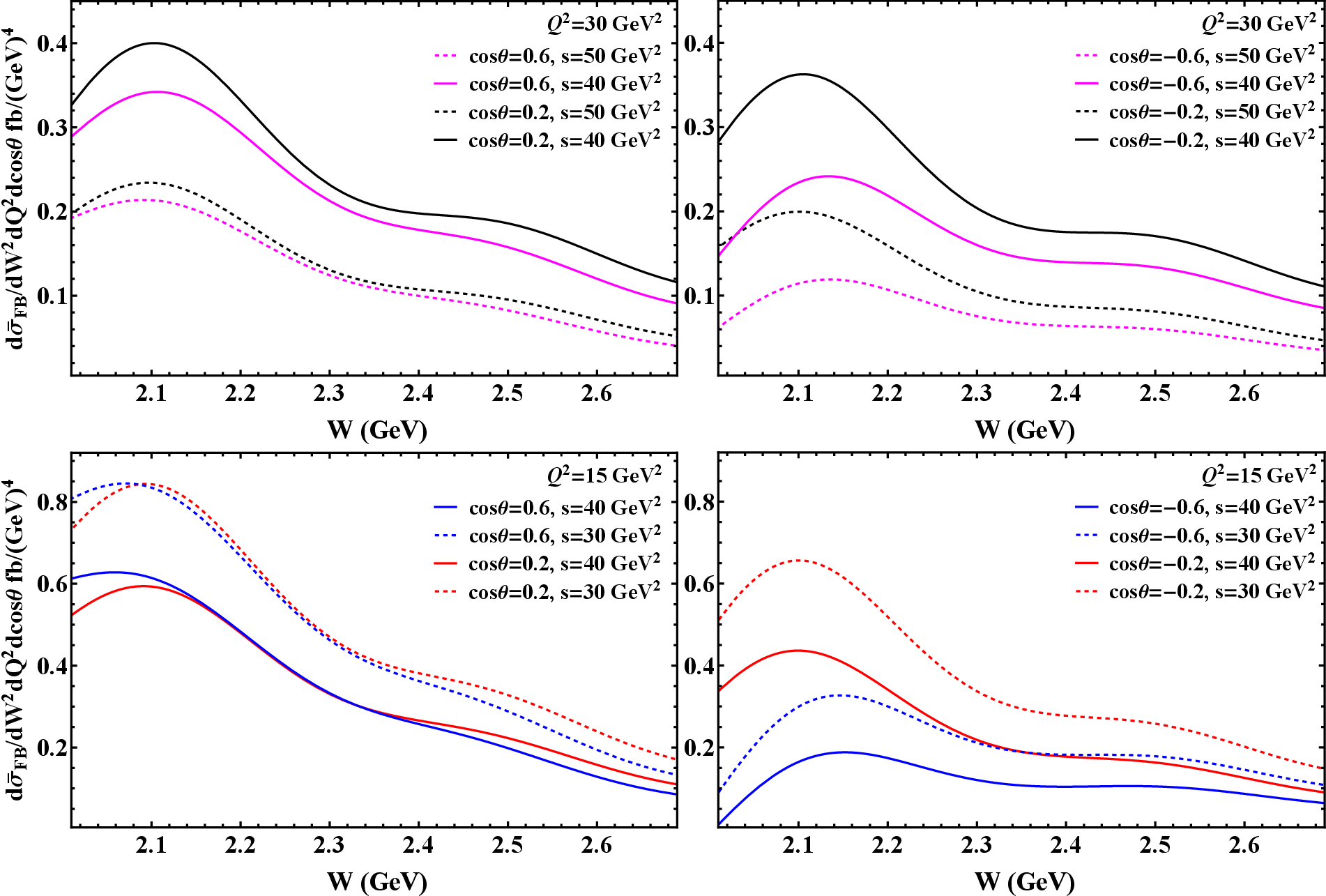}
\caption{ The cross section difference (Eq.\ref{intasym}) allows to measure the interference term of $e^- \gamma  \to e^- p \bar{p}  $. It depends on the invariant mass of the proton-antiproton pair and on the polar angle $\theta$ of the nucleon.}
\label{fig:crossint}
\end{figure}

For the interference term of two subprocesses, we use  Eq.~\eqref{intasym} instead of Eq.~\eqref{crointph} for numerical estimates since more GDAs  can be extracted from the angular dependence in this forward-backward asymmetry.
The differential cross section of 
Eq.~\eqref{intasym} is depicted in Fig.~\ref{fig:crossint}. In the top panel, we take $\cos \theta=\pm0.2$ ($\pm0.6$) for the black (magenta) curves, with $Q^2 = 30~\text{GeV}^2$ fixed.
The  solid and dashed  lines correspond to $s =40~\text{GeV}^2$ and $s =50~\text{GeV}^2$, respectively.  The magnitude of the interference term is comparable to that of the pure QCD contribution at $Q^2 = 30~\text{GeV}^2$, as shown in Fig.~\ref{fig:crossg}. In the bottom panel, $s = 30~\text{GeV}^2$ ($40~\text{GeV}^2$) is fixed for the dashed (solid) curves at $Q^2 = 15~\text{GeV}^2$, with the red and blue curves corresponding to $\cos\theta = \pm 0.2$ and $\cos\theta = \pm 0.6$, respectively. An enhancement of the interference term is  observed when $Q^2$ decreases from $30~\text{GeV}^2$ to $15~\text{GeV}^2$, which also arises from the prefactor in Eq.~\eqref{intasym}.
Our numerical results can only be regarded as order-of-magnitude estimates for the interference term and the pure QCD contribution, since the proton–antiproton GDAs are poorly known. Further theoretical  studies of  GDAs are necessary to obtain  precise predictions, and precise experimental analysis of the reaction is mostly needed.

\begin{figure}[htb]
\centering
\includegraphics[width=0.9\textwidth]{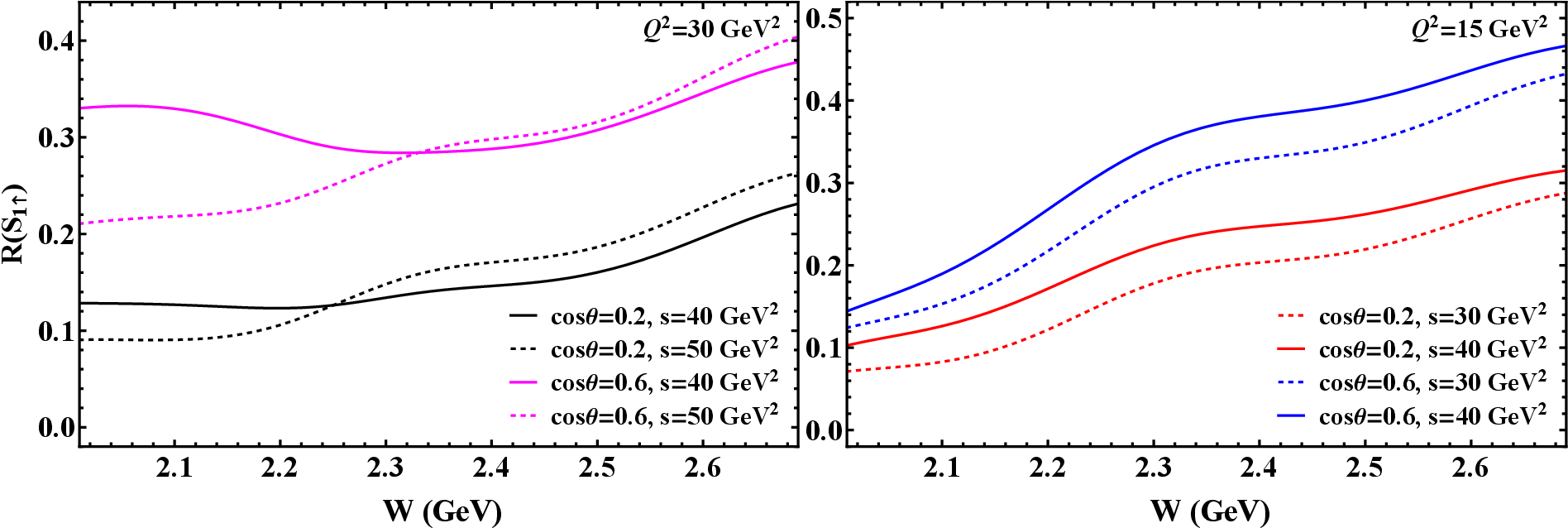}
\caption{ The estimate of ratio  $R(S_{1 \uparrow})$ in $e^- \gamma  \to  e^-  p \bar{p} $ using models of the proton EM FFs and Compton FFs.}
\label{fig:ratio}
\end{figure}

After integrating over the azimuthal angle, one can see that the dependence on $S_1^x$ and $S_1^z$ disappears in the single-spin correlation. Thus,  the y-axis is chosen as the polarization direction in the rest frame of the baryon. We can express the spin vector of the spin-up baryon as
\begin{equation}
S_{1 \uparrow}^{\mu}=(0, 0, 1, 0),
\label{eqn:spin-up}
\end{equation}
and $S_{1 \downarrow}^{\mu} =-S_{1 \uparrow}^{\mu} $ for the spin-down state. The cross sections for the production of spin-up and spin-down baryons are  given respectively by
\begin{equation}
\begin{aligned}
d \sigma(S_{1 \uparrow})& =\frac{1}{2}d \bar{\sigma}  + d \hat{\sigma}(S_{1 \uparrow}), \\
d \sigma(S_{1 \downarrow})& =\frac{1}{2}d \bar{\sigma} - d \hat{\sigma}(S_{1 \uparrow}).
\label{eqn:up-down}
\end{aligned}
\end{equation}
To see the polarization effect, one can define the following ratio using the cross sections derived in 
Sec.~\ref{sacs},
\begin{equation}
R(S_{1 \uparrow})= \left[ \frac{d \hat{\sigma}(S_{1 \uparrow}) }{d\hat{s}dQ^{2} d\left ( \cos \theta  \right )    } \right] \Big /
\left[ \frac{1}{2}
\frac{d \bar{\sigma}  }{d\hat{s}dQ^{2} d\left ( \cos \theta  \right )   } \right].
\label{eqn:ratio}
\end{equation}
The single-spin correlation $d\hat{\sigma}$ is proportional to  the terms  $\mathrm{Im}\left ( F_{i}F_{j}^{\ast }     \right )$, where $F_i$ represents a EM FF or Compton FF as indicated by Eqs.~\eqref{gdaspin},  \eqref{emspin}, and \eqref{sscin}. 
However, the phases of these FFs, $F_i = |F_i| e^{i\delta_i}$, are still unknown, and there is no reason to assume they vanish. In this work, we show the maximal single-spin correlation by setting all $\sin(\delta_i - \delta_j) = 1$ in $\mathrm{Im}\left(F_i F_j^{\ast}\right)$. The numerical estimate of the ratio $R(S_{1 \uparrow})$ is shown in Fig.~\ref{fig:ratio}, based on models of the proton EM FFs and Compton FFs. We choose typical values for 
$s$  and $Q^2$ according to the kinematics of Belle II.
The colors of curves indicate the different $\cos \theta$, with the same conventions as  Fig.~\ref{fig:crob}. We can infer that 
the polarization effect is sizable in the cross section.

\section{Summary}
\label{summary}
Hadronic GPDs and GDAs are three-dimensional functions that characterize the internal structure of hadrons. The second moments of GPDs and GDAs give rise to the GFFs of hadrons in the spacelike and timelike regions, respectively, which  reveal  key information about hadron structure.
As most hadrons are not stable, it is difficult to access their GPDs using  exclusive processes. However, hadronic GDAs probed through hadron–pair production processes such as $\gamma^{\ast} \gamma \to B \bar{B}$ and $\gamma^{\ast} \to B \bar{B} \gamma$ provide access not only to nucleons but also to unstable baryons such as $\Lambda$ and $\Sigma$.

Recently, the Belle Collaboration reported the cross section for $\gamma^{\ast} \gamma \to \pi^0 \pi^0$, where the photons are emitted from the $e^{\pm}$ beams~\cite{Belle:2015oin}. The pion GFFs were subsequently extracted from these experimental data for the first time~\cite{Kumano:2017lhr}. In this work, we extend the GDA study  to a spin-1/2 baryon-antibaryon pair, namely $ e^{\pm} \gamma \to e^{\pm} B \bar{B}$, which can be measured at Belle II, the proposed STCF, and electron–ion  colliders. In this process, the baryon GDAs are involved in the two-photon fusion subprocess with the $C$-even baryon-antibaryon pairs. In addition, there is the bremsstrahlung subprocess, and  the $C$-odd baryon-antibaryon pairs are produced. We calculate the theoretical cross sections for $ e^{\pm} \gamma \to e^{\pm} B \bar{B}$.
The interference contribution of two subprocesses can be extracted using the lepton charge asymmetry, which are expressed in terms of the baryon GDAs and EM FFs. Since one can determine the spin vectors of baryons from their decays such as $\Lambda \to N \pi$ and $\Sigma \to N \pi$, we also include the spin vector of the baryon (or antibaryon), and define  the single-spin correlation, from which additional information on the baryon GDAs can be probed. We use motivated models of EM FFs and GDAs to present the 
 numerical estimates for the process $ e^- \gamma \to e^- p \bar{p}$, where the kinematics are chosen according to the Belle II experiment. 
 Our results show that we can find a kinematical region where 
 the GDA contribution may dominate  the cross section, which can be used for the experimental measurements of this process. Our study thus provides useful guidance for the future extraction of baryon GDAs at Belle II.

\section*{Acknowledgments}
 Qin-Tao Song was supported by the National Natural Science Foundation
of China under Grant No. 12005191 and by the Natural Science Foundation of Henan Province under Grant  No. 252300423011.

\bibliography{bibliography}

\begin{thebibliography}{76}
\expandafter\ifx\csname natexlab\endcsname\relax\def\natexlab#1{#1}\fi
\expandafter\ifx\csname bibnamefont\endcsname\relax
  \def\bibnamefont#1{#1}\fi
\expandafter\ifx\csname bibfnamefont\endcsname\relax
  \def\bibfnamefont#1{#1}\fi
\expandafter\ifx\csname citenamefont\endcsname\relax
  \def\citenamefont#1{#1}\fi
\expandafter\ifx\csname url\endcsname\relax
  \def\url#1{\texttt{#1}}\fi
\expandafter\ifx\csname urlprefix\endcsname\relax\def\urlprefix{URL }\fi
\providecommand{\bibinfo}[2]{#2}
\providecommand{\eprint}[2][]{\url{#2}}

\bibitem[{\citenamefont{M{\"u}ller et~al.}(1994)\citenamefont{M{\"u}ller,
  Robaschik, Geyer, Dittes, and Ho{\v{r}}ej{\v{s}}i}}]{Muller:1994ses}
\bibinfo{author}{\bibfnamefont{D.}~\bibnamefont{M{\"u}ller}},
  \bibinfo{author}{\bibfnamefont{D.}~\bibnamefont{Robaschik}},
  \bibinfo{author}{\bibfnamefont{B.}~\bibnamefont{Geyer}},
  \bibinfo{author}{\bibfnamefont{F.~M.} \bibnamefont{Dittes}},
  \bibnamefont{and}
  \bibinfo{author}{\bibfnamefont{J.}~\bibnamefont{Ho{\v{r}}ej{\v{s}}i}},
  \bibinfo{journal}{Fortsch. Phys.} \textbf{\bibinfo{volume}{42}},
  \bibinfo{pages}{101} (\bibinfo{year}{1994}), \eprint{hep-ph/9812448}.

\bibitem[{\citenamefont{Diehl et~al.}(1998)\citenamefont{Diehl, Gousset, Pire,
  and Teryaev}}]{Diehl:1998dk}
\bibinfo{author}{\bibfnamefont{M.}~\bibnamefont{Diehl}},
  \bibinfo{author}{\bibfnamefont{T.}~\bibnamefont{Gousset}},
  \bibinfo{author}{\bibfnamefont{B.}~\bibnamefont{Pire}}, \bibnamefont{and}
  \bibinfo{author}{\bibfnamefont{O.}~\bibnamefont{Teryaev}},
  \bibinfo{journal}{Phys. Rev. Lett.} \textbf{\bibinfo{volume}{81}},
  \bibinfo{pages}{1782} (\bibinfo{year}{1998}), \eprint{hep-ph/9805380}.

\bibitem[{\citenamefont{Polyakov}(1999)}]{Polyakov:1998ze}
\bibinfo{author}{\bibfnamefont{M.~V.} \bibnamefont{Polyakov}},
  \bibinfo{journal}{Nucl. Phys. B} \textbf{\bibinfo{volume}{555}},
  \bibinfo{pages}{231} (\bibinfo{year}{1999}), \eprint{hep-ph/9809483}.

\bibitem[{\citenamefont{Ji}(1997)}]{Ji:1996ek}
\bibinfo{author}{\bibfnamefont{X.-D.} \bibnamefont{Ji}},
  \bibinfo{journal}{Phys. Rev. Lett.} \textbf{\bibinfo{volume}{78}},
  \bibinfo{pages}{610} (\bibinfo{year}{1997}), \eprint{hep-ph/9603249}.

\bibitem[{\citenamefont{Radyushkin}(1999)}]{Radyushkin:1998bz}
\bibinfo{author}{\bibfnamefont{A.~V.} \bibnamefont{Radyushkin}},
  \bibinfo{journal}{Phys. Lett. B} \textbf{\bibinfo{volume}{449}},
  \bibinfo{pages}{81} (\bibinfo{year}{1999}), \eprint{hep-ph/9810466}.

\bibitem[{\citenamefont{Burkardt}(2000)}]{Burkardt:2000za}
\bibinfo{author}{\bibfnamefont{M.}~\bibnamefont{Burkardt}},
  \bibinfo{journal}{Phys. Rev. D} \textbf{\bibinfo{volume}{62}},
  \bibinfo{pages}{071503} (\bibinfo{year}{2000}), \bibinfo{note}{[Erratum:
  Phys.Rev.D 66, 119903 (2002)]}, \eprint{hep-ph/0005108}.

\bibitem[{\citenamefont{Ralston and Pire}(2002)}]{Ralston:2001xs}
\bibinfo{author}{\bibfnamefont{J.~P.} \bibnamefont{Ralston}} \bibnamefont{and}
  \bibinfo{author}{\bibfnamefont{B.}~\bibnamefont{Pire}},
  \bibinfo{journal}{Phys. Rev. D} \textbf{\bibinfo{volume}{66}},
  \bibinfo{pages}{111501} (\bibinfo{year}{2002}), \eprint{hep-ph/0110075}.

\bibitem[{\citenamefont{Diehl}(2002)}]{Diehl:2002he}
\bibinfo{author}{\bibfnamefont{M.}~\bibnamefont{Diehl}}, \bibinfo{journal}{Eur.
  Phys. J. C} \textbf{\bibinfo{volume}{25}}, \bibinfo{pages}{223}
  (\bibinfo{year}{2002}), \bibinfo{note}{[Erratum: Eur.Phys.J.C 31, 277--278
  (2003)]}, \eprint{hep-ph/0205208}.

\bibitem[{\citenamefont{Pire and Szymanowski}(2003)}]{Pire:2002ut}
\bibinfo{author}{\bibfnamefont{B.}~\bibnamefont{Pire}} \bibnamefont{and}
  \bibinfo{author}{\bibfnamefont{L.}~\bibnamefont{Szymanowski}},
  \bibinfo{journal}{Phys. Lett. B} \textbf{\bibinfo{volume}{556}},
  \bibinfo{pages}{129} (\bibinfo{year}{2003}), \eprint{hep-ph/0212296}.

\bibitem[{\citenamefont{Leader and Lorc\'e}(2014)}]{Leader:2013jra}
\bibinfo{author}{\bibfnamefont{E.}~\bibnamefont{Leader}} \bibnamefont{and}
  \bibinfo{author}{\bibfnamefont{C.}~\bibnamefont{Lorc\'e}},
  \bibinfo{journal}{Phys. Rept.} \textbf{\bibinfo{volume}{541}},
  \bibinfo{pages}{163} (\bibinfo{year}{2014}), \eprint{1309.4235}.

\bibitem[{\citenamefont{Ji et~al.}(2021)\citenamefont{Ji, Yuan, and
  Zhao}}]{Ji:2020ena}
\bibinfo{author}{\bibfnamefont{X.}~\bibnamefont{Ji}},
  \bibinfo{author}{\bibfnamefont{F.}~\bibnamefont{Yuan}}, \bibnamefont{and}
  \bibinfo{author}{\bibfnamefont{Y.}~\bibnamefont{Zhao}},
  \bibinfo{journal}{Nature Rev. Phys.} \textbf{\bibinfo{volume}{3}},
  \bibinfo{pages}{27} (\bibinfo{year}{2021}), \eprint{2009.01291}.

\bibitem[{\citenamefont{Aidala et~al.}(2013)\citenamefont{Aidala, Bass, Hasch,
  and Mallot}}]{Aidala:2012mv}
\bibinfo{author}{\bibfnamefont{C.~A.} \bibnamefont{Aidala}},
  \bibinfo{author}{\bibfnamefont{S.~D.} \bibnamefont{Bass}},
  \bibinfo{author}{\bibfnamefont{D.}~\bibnamefont{Hasch}}, \bibnamefont{and}
  \bibinfo{author}{\bibfnamefont{G.~K.} \bibnamefont{Mallot}},
  \bibinfo{journal}{Rev. Mod. Phys.} \textbf{\bibinfo{volume}{85}},
  \bibinfo{pages}{655} (\bibinfo{year}{2013}), \eprint{1209.2803}.

\bibitem[{\citenamefont{Polyakov}(2003)}]{Polyakov:2002yz}
\bibinfo{author}{\bibfnamefont{M.~V.} \bibnamefont{Polyakov}},
  \bibinfo{journal}{Phys. Lett. B} \textbf{\bibinfo{volume}{555}},
  \bibinfo{pages}{57} (\bibinfo{year}{2003}), \eprint{hep-ph/0210165}.

\bibitem[{\citenamefont{Polyakov and Schweitzer}(2018)}]{Polyakov:2018zvc}
\bibinfo{author}{\bibfnamefont{M.~V.} \bibnamefont{Polyakov}} \bibnamefont{and}
  \bibinfo{author}{\bibfnamefont{P.}~\bibnamefont{Schweitzer}},
  \bibinfo{journal}{Int. J. Mod. Phys. A} \textbf{\bibinfo{volume}{33}},
  \bibinfo{pages}{1830025} (\bibinfo{year}{2018}), \eprint{1805.06596}.

\bibitem[{\citenamefont{Burkert et~al.}(2018)\citenamefont{Burkert,
  Elouadrhiri, and Girod}}]{Burkert:2018bqq}
\bibinfo{author}{\bibfnamefont{V.~D.} \bibnamefont{Burkert}},
  \bibinfo{author}{\bibfnamefont{L.}~\bibnamefont{Elouadrhiri}},
  \bibnamefont{and} \bibinfo{author}{\bibfnamefont{F.~X.} \bibnamefont{Girod}},
  \bibinfo{journal}{Nature} \textbf{\bibinfo{volume}{557}},
  \bibinfo{pages}{396} (\bibinfo{year}{2018}).

\bibitem[{\citenamefont{Lorc\'e et~al.}(2019)\citenamefont{Lorc\'e, Moutarde,
  and Trawi\'nski}}]{Lorce:2018egm}
\bibinfo{author}{\bibfnamefont{C.}~\bibnamefont{Lorc\'e}},
  \bibinfo{author}{\bibfnamefont{H.}~\bibnamefont{Moutarde}}, \bibnamefont{and}
  \bibinfo{author}{\bibfnamefont{A.~P.} \bibnamefont{Trawi\'nski}},
  \bibinfo{journal}{Eur. Phys. J. C} \textbf{\bibinfo{volume}{79}},
  \bibinfo{pages}{89} (\bibinfo{year}{2019}), \eprint{1810.09837}.

\bibitem[{\citenamefont{Kumeri\v{c}ki}(2019)}]{Kumericki:2019ddg}
\bibinfo{author}{\bibfnamefont{K.}~\bibnamefont{Kumeri\v{c}ki}},
  \bibinfo{journal}{Nature} \textbf{\bibinfo{volume}{570}}, \bibinfo{pages}{E1}
  (\bibinfo{year}{2019}).

\bibitem[{\citenamefont{Burkert et~al.}(2023)\citenamefont{Burkert,
  Elouadrhiri, Girod, Lorc\'e, Schweitzer, and Shanahan}}]{Burkert:2023wzr}
\bibinfo{author}{\bibfnamefont{V.~D.} \bibnamefont{Burkert}},
  \bibinfo{author}{\bibfnamefont{L.}~\bibnamefont{Elouadrhiri}},
  \bibinfo{author}{\bibfnamefont{F.~X.} \bibnamefont{Girod}},
  \bibinfo{author}{\bibfnamefont{C.}~\bibnamefont{Lorc\'e}},
  \bibinfo{author}{\bibfnamefont{P.}~\bibnamefont{Schweitzer}},
  \bibnamefont{and} \bibinfo{author}{\bibfnamefont{P.~E.}
  \bibnamefont{Shanahan}}, \bibinfo{journal}{Rev. Mod. Phys.}
  \textbf{\bibinfo{volume}{95}}, \bibinfo{pages}{041002}
  (\bibinfo{year}{2023}), \eprint{2303.08347}.

\bibitem[{\citenamefont{Freese and Miller}(2021)}]{Freese:2021qtb}
\bibinfo{author}{\bibfnamefont{A.}~\bibnamefont{Freese}} \bibnamefont{and}
  \bibinfo{author}{\bibfnamefont{G.~A.} \bibnamefont{Miller}},
  \bibinfo{journal}{Phys. Rev. D} \textbf{\bibinfo{volume}{104}},
  \bibinfo{pages}{014024} (\bibinfo{year}{2021}), \eprint{2104.03213}.

\bibitem[{\citenamefont{Fujii and Tanaka}(2025)}]{Fujii:2025pkv}
\bibinfo{author}{\bibfnamefont{D.}~\bibnamefont{Fujii}} \bibnamefont{and}
  \bibinfo{author}{\bibfnamefont{M.}~\bibnamefont{Tanaka}},
  \bibinfo{journal}{Phys. Lett. B} \textbf{\bibinfo{volume}{870}},
  \bibinfo{pages}{139872} (\bibinfo{year}{2025}).

\bibitem[{\citenamefont{Dutrieux et~al.}(2025)\citenamefont{Dutrieux, Meisgny,
  Mezrag, and Moutarde}}]{Dutrieux:2024bgc}
\bibinfo{author}{\bibfnamefont{H.}~\bibnamefont{Dutrieux}},
  \bibinfo{author}{\bibfnamefont{T.}~\bibnamefont{Meisgny}},
  \bibinfo{author}{\bibfnamefont{C.}~\bibnamefont{Mezrag}}, \bibnamefont{and}
  \bibinfo{author}{\bibfnamefont{H.}~\bibnamefont{Moutarde}},
  \bibinfo{journal}{Eur. Phys. J. C} \textbf{\bibinfo{volume}{85}},
  \bibinfo{pages}{105} (\bibinfo{year}{2025}), \eprint{2410.13518}.

\bibitem[{\citenamefont{Garc\'\i{}a Mart\'\i{}n-Caro
  et~al.}(2024)\citenamefont{Garc\'\i{}a Mart\'\i{}n-Caro, Huidobro, and
  Hatta}}]{GarciaMartin-Caro:2023toa}
\bibinfo{author}{\bibfnamefont{A.}~\bibnamefont{Garc\'\i{}a Mart\'\i{}n-Caro}},
  \bibinfo{author}{\bibfnamefont{M.}~\bibnamefont{Huidobro}}, \bibnamefont{and}
  \bibinfo{author}{\bibfnamefont{Y.}~\bibnamefont{Hatta}},
  \bibinfo{journal}{Phys. Rev. D} \textbf{\bibinfo{volume}{110}},
  \bibinfo{pages}{034002} (\bibinfo{year}{2024}), \eprint{2312.12984}.

\bibitem[{\citenamefont{Li and Vary}(2024)}]{Li:2023izn}
\bibinfo{author}{\bibfnamefont{Y.}~\bibnamefont{Li}} \bibnamefont{and}
  \bibinfo{author}{\bibfnamefont{J.~P.} \bibnamefont{Vary}},
  \bibinfo{journal}{Phys. Rev. D} \textbf{\bibinfo{volume}{109}},
  \bibinfo{pages}{L051501} (\bibinfo{year}{2024}), \eprint{2312.02543}.

\bibitem[{\citenamefont{Lorc\'e and Song}(2025)}]{Lorce:2025ayr}
\bibinfo{author}{\bibfnamefont{C.}~\bibnamefont{Lorc\'e}} \bibnamefont{and}
  \bibinfo{author}{\bibfnamefont{Q.-T.} \bibnamefont{Song}},
  \bibinfo{journal}{Phys. Lett. B} \textbf{\bibinfo{volume}{864}},
  \bibinfo{pages}{139433} (\bibinfo{year}{2025}), \eprint{2501.05092}.

\bibitem[{\citenamefont{Hu et~al.}(2025)\citenamefont{Hu, Cao, Xu, Li, Zhao,
  and Vary}}]{Hu:2024edc}
\bibinfo{author}{\bibfnamefont{T.}~\bibnamefont{Hu}},
  \bibinfo{author}{\bibfnamefont{X.}~\bibnamefont{Cao}},
  \bibinfo{author}{\bibfnamefont{S.}~\bibnamefont{Xu}},
  \bibinfo{author}{\bibfnamefont{Y.}~\bibnamefont{Li}},
  \bibinfo{author}{\bibfnamefont{X.}~\bibnamefont{Zhao}}, \bibnamefont{and}
  \bibinfo{author}{\bibfnamefont{J.~P.} \bibnamefont{Vary}},
  \bibinfo{journal}{Phys. Rev. D} \textbf{\bibinfo{volume}{111}},
  \bibinfo{pages}{074031} (\bibinfo{year}{2025}), \eprint{2408.09689}.

\bibitem[{\citenamefont{Broniowski and
  Ruiz~Arriola}(2025)}]{Broniowski:2025ctl}
\bibinfo{author}{\bibfnamefont{W.}~\bibnamefont{Broniowski}} \bibnamefont{and}
  \bibinfo{author}{\bibfnamefont{E.}~\bibnamefont{Ruiz~Arriola}},
  \bibinfo{journal}{Phys. Rev. D} \textbf{\bibinfo{volume}{112}},
  \bibinfo{pages}{054028} (\bibinfo{year}{2025}), \eprint{2503.09297}.

\bibitem[{\citenamefont{Diehl et~al.}(2000)\citenamefont{Diehl, Gousset, and
  Pire}}]{Diehl:2000uv}
\bibinfo{author}{\bibfnamefont{M.}~\bibnamefont{Diehl}},
  \bibinfo{author}{\bibfnamefont{T.}~\bibnamefont{Gousset}}, \bibnamefont{and}
  \bibinfo{author}{\bibfnamefont{B.}~\bibnamefont{Pire}},
  \bibinfo{journal}{Phys. Rev. D} \textbf{\bibinfo{volume}{62}},
  \bibinfo{pages}{073014} (\bibinfo{year}{2000}), \eprint{hep-ph/0003233}.

\bibitem[{\citenamefont{Kivel et~al.}(1999)\citenamefont{Kivel, Mankiewicz, and
  Polyakov}}]{Kivel:1999sd}
\bibinfo{author}{\bibfnamefont{N.}~\bibnamefont{Kivel}},
  \bibinfo{author}{\bibfnamefont{L.}~\bibnamefont{Mankiewicz}},
  \bibnamefont{and} \bibinfo{author}{\bibfnamefont{M.~V.}
  \bibnamefont{Polyakov}}, \bibinfo{journal}{Phys. Lett. B}
  \textbf{\bibinfo{volume}{467}}, \bibinfo{pages}{263} (\bibinfo{year}{1999}),
  \eprint{hep-ph/9908334}.

\bibitem[{\citenamefont{Kumano et~al.}(2018)\citenamefont{Kumano, Song, and
  Teryaev}}]{Kumano:2017lhr}
\bibinfo{author}{\bibfnamefont{S.}~\bibnamefont{Kumano}},
  \bibinfo{author}{\bibfnamefont{Q.-T.} \bibnamefont{Song}}, \bibnamefont{and}
  \bibinfo{author}{\bibfnamefont{O.~V.} \bibnamefont{Teryaev}},
  \bibinfo{journal}{Phys. Rev. D} \textbf{\bibinfo{volume}{97}},
  \bibinfo{pages}{014020} (\bibinfo{year}{2018}), \eprint{1711.08088}.

\bibitem[{\citenamefont{Lorc\'e
  et~al.}(2022{\natexlab{a}})\citenamefont{Lorc\'e, Pire, and
  Song}}]{Lorce:2022tiq}
\bibinfo{author}{\bibfnamefont{C.}~\bibnamefont{Lorc\'e}},
  \bibinfo{author}{\bibfnamefont{B.}~\bibnamefont{Pire}}, \bibnamefont{and}
  \bibinfo{author}{\bibfnamefont{Q.-T.} \bibnamefont{Song}},
  \bibinfo{journal}{Phys. Rev. D} \textbf{\bibinfo{volume}{106}},
  \bibinfo{pages}{094030} (\bibinfo{year}{2022}{\natexlab{a}}),
  \eprint{2209.11140}.

\bibitem[{\citenamefont{Lorc\'e
  et~al.}(2022{\natexlab{b}})\citenamefont{Lorc\'e, Pire, and
  Song}}]{Lorce:2022cze}
\bibinfo{author}{\bibfnamefont{C.}~\bibnamefont{Lorc\'e}},
  \bibinfo{author}{\bibfnamefont{B.}~\bibnamefont{Pire}}, \bibnamefont{and}
  \bibinfo{author}{\bibfnamefont{Q.-T.} \bibnamefont{Song}}, in
  \emph{\bibinfo{booktitle}{{29th International Workshop on Deep-Inelastic
  Scattering and Related Subjects}}} (\bibinfo{year}{2022}{\natexlab{b}}),
  \eprint{2208.12532}.

\bibitem[{\citenamefont{Song et~al.}(2025)\citenamefont{Song, Teryaev, and
  Yoshida}}]{Song:2025zwl}
\bibinfo{author}{\bibfnamefont{Q.-T.} \bibnamefont{Song}},
  \bibinfo{author}{\bibfnamefont{O.~V.} \bibnamefont{Teryaev}},
  \bibnamefont{and} \bibinfo{author}{\bibfnamefont{S.}~\bibnamefont{Yoshida}},
  \bibinfo{journal}{Phys. Lett. B} \textbf{\bibinfo{volume}{868}},
  \bibinfo{pages}{139797} (\bibinfo{year}{2025}), \eprint{2503.11316}.

\bibitem[{\citenamefont{Lu and Schmidt}(2006)}]{Lu:2006ut}
\bibinfo{author}{\bibfnamefont{Z.}~\bibnamefont{Lu}} \bibnamefont{and}
  \bibinfo{author}{\bibfnamefont{I.}~\bibnamefont{Schmidt}},
  \bibinfo{journal}{Phys. Rev. D} \textbf{\bibinfo{volume}{73}},
  \bibinfo{pages}{094021} (\bibinfo{year}{2006}), \bibinfo{note}{[Erratum:
  Phys.Rev.D 75, 099902 (2007)]}, \eprint{hep-ph/0603151}.

\bibitem[{\citenamefont{Pire and Song}(2023)}]{Pire:2023kng}
\bibinfo{author}{\bibfnamefont{B.}~\bibnamefont{Pire}} \bibnamefont{and}
  \bibinfo{author}{\bibfnamefont{Q.-T.} \bibnamefont{Song}},
  \bibinfo{journal}{Phys. Rev. D} \textbf{\bibinfo{volume}{107}},
  \bibinfo{pages}{114014} (\bibinfo{year}{2023}), \eprint{2304.06389}.

\bibitem[{\citenamefont{Pire and Song}(2024)}]{Pire:2023ztb}
\bibinfo{author}{\bibfnamefont{B.}~\bibnamefont{Pire}} \bibnamefont{and}
  \bibinfo{author}{\bibfnamefont{Q.-T.} \bibnamefont{Song}},
  \bibinfo{journal}{Phys. Rev. D} \textbf{\bibinfo{volume}{109}},
  \bibinfo{pages}{074016} (\bibinfo{year}{2024}), \eprint{2311.06005}.

\bibitem[{\citenamefont{Han et~al.}(2025)\citenamefont{Han, Pire, and
  Song}}]{Han:2025mvq}
\bibinfo{author}{\bibfnamefont{J.}~\bibnamefont{Han}},
  \bibinfo{author}{\bibfnamefont{B.}~\bibnamefont{Pire}}, \bibnamefont{and}
  \bibinfo{author}{\bibfnamefont{Q.-T.} \bibnamefont{Song}},
  \bibinfo{journal}{Phys. Rev. D} \textbf{\bibinfo{volume}{112}},
  \bibinfo{pages}{014048} (\bibinfo{year}{2025}), \eprint{2506.09854}.

\bibitem[{\citenamefont{Bhattacharya et~al.}(2025)\citenamefont{Bhattacharya,
  Boussarie, Pire, and Szymanowski}}]{Bhattacharya:2025awq}
\bibinfo{author}{\bibfnamefont{S.}~\bibnamefont{Bhattacharya}},
  \bibinfo{author}{\bibfnamefont{R.}~\bibnamefont{Boussarie}},
  \bibinfo{author}{\bibfnamefont{B.}~\bibnamefont{Pire}}, \bibnamefont{and}
  \bibinfo{author}{\bibfnamefont{L.}~\bibnamefont{Szymanowski}}
  (\bibinfo{year}{2025}), \eprint{2507.23529}.

\bibitem[{\citenamefont{Masuda et~al.}(2016)}]{Belle:2015oin}
\bibinfo{author}{\bibfnamefont{M.}~\bibnamefont{Masuda}} \bibnamefont{et~al.}
  (\bibinfo{collaboration}{Belle}), \bibinfo{journal}{Phys. Rev. D}
  \textbf{\bibinfo{volume}{93}}, \bibinfo{pages}{032003}
  (\bibinfo{year}{2016}), \eprint{1508.06757}.

\bibitem[{\citenamefont{Achasov et~al.}(2024)}]{Achasov:2023gey}
\bibinfo{author}{\bibfnamefont{M.}~\bibnamefont{Achasov}} \bibnamefont{et~al.},
  \bibinfo{journal}{Front. Phys. (Beijing)} \textbf{\bibinfo{volume}{19}},
  \bibinfo{pages}{14701} (\bibinfo{year}{2024}), \eprint{2303.15790}.

\bibitem[{\citenamefont{Akhmetshin et~al.}(2019)}]{CMD-3:2018kql}
\bibinfo{author}{\bibfnamefont{R.~R.} \bibnamefont{Akhmetshin}}
  \bibnamefont{et~al.} (\bibinfo{collaboration}{CMD-3}),
  \bibinfo{journal}{Phys. Lett. B} \textbf{\bibinfo{volume}{794}},
  \bibinfo{pages}{64} (\bibinfo{year}{2019}), \eprint{1808.00145}.

\bibitem[{\citenamefont{Ablikim et~al.}(2021{\natexlab{a}})}]{BESIII:2021tbq}
\bibinfo{author}{\bibfnamefont{M.}~\bibnamefont{Ablikim}} \bibnamefont{et~al.}
  (\bibinfo{collaboration}{BESIII}), \bibinfo{journal}{Nature Phys.}
  \textbf{\bibinfo{volume}{17}}, \bibinfo{pages}{1200}
  (\bibinfo{year}{2021}{\natexlab{a}}), \eprint{2103.12486}.

\bibitem[{\citenamefont{Lees et~al.}(2013)}]{BaBar:2013ves}
\bibinfo{author}{\bibfnamefont{J.~P.} \bibnamefont{Lees}} \bibnamefont{et~al.}
  (\bibinfo{collaboration}{BaBar}), \bibinfo{journal}{Phys. Rev. D}
  \textbf{\bibinfo{volume}{87}}, \bibinfo{pages}{092005}
  (\bibinfo{year}{2013}), \eprint{1302.0055}.

\bibitem[{\citenamefont{Ablikim et~al.}(2021{\natexlab{b}})}]{BESIII:2021rqk}
\bibinfo{author}{\bibfnamefont{M.}~\bibnamefont{Ablikim}} \bibnamefont{et~al.}
  (\bibinfo{collaboration}{BESIII}), \bibinfo{journal}{Phys. Lett. B}
  \textbf{\bibinfo{volume}{817}}, \bibinfo{pages}{136328}
  (\bibinfo{year}{2021}{\natexlab{b}}), \eprint{2102.10337}.

\bibitem[{\citenamefont{Ablikim et~al.}(2020)}]{BESIII:2019hdp}
\bibinfo{author}{\bibfnamefont{M.}~\bibnamefont{Ablikim}} \bibnamefont{et~al.}
  (\bibinfo{collaboration}{BESIII}), \bibinfo{journal}{Phys. Rev. Lett.}
  \textbf{\bibinfo{volume}{124}}, \bibinfo{pages}{042001}
  (\bibinfo{year}{2020}), \eprint{1905.09001}.

\bibitem[{\citenamefont{Ablikim et~al.}(2023{\natexlab{a}})}]{BESIII:2022rrg}
\bibinfo{author}{\bibfnamefont{M.}~\bibnamefont{Ablikim}} \bibnamefont{et~al.}
  (\bibinfo{collaboration}{BESIII}), \bibinfo{journal}{Phys. Rev. Lett.}
  \textbf{\bibinfo{volume}{130}}, \bibinfo{pages}{151905}
  (\bibinfo{year}{2023}{\natexlab{a}}), \eprint{2212.07071}.

\bibitem[{\citenamefont{Ablikim et~al.}(2019{\natexlab{a}})}]{BESIII:2019tgo}
\bibinfo{author}{\bibfnamefont{M.}~\bibnamefont{Ablikim}} \bibnamefont{et~al.}
  (\bibinfo{collaboration}{BESIII}), \bibinfo{journal}{Phys. Rev. D}
  \textbf{\bibinfo{volume}{99}}, \bibinfo{pages}{092002}
  (\bibinfo{year}{2019}{\natexlab{a}}), \eprint{1902.00665}.

\bibitem[{\citenamefont{Ablikim et~al.}(2018)}]{BESIII:2017hyw}
\bibinfo{author}{\bibfnamefont{M.}~\bibnamefont{Ablikim}} \bibnamefont{et~al.}
  (\bibinfo{collaboration}{BESIII}), \bibinfo{journal}{Phys. Rev. D}
  \textbf{\bibinfo{volume}{97}}, \bibinfo{pages}{032013}
  (\bibinfo{year}{2018}), \eprint{1709.10236}.

\bibitem[{\citenamefont{Ablikim et~al.}(2019{\natexlab{b}})}]{BESIII:2019nep}
\bibinfo{author}{\bibfnamefont{M.}~\bibnamefont{Ablikim}} \bibnamefont{et~al.}
  (\bibinfo{collaboration}{BESIII}), \bibinfo{journal}{Phys. Rev. Lett.}
  \textbf{\bibinfo{volume}{123}}, \bibinfo{pages}{122003}
  (\bibinfo{year}{2019}{\natexlab{b}}), \eprint{1903.09421}.

\bibitem[{\citenamefont{Ablikim et~al.}(2023{\natexlab{b}})}]{BESIII:2023ioy}
\bibinfo{author}{\bibfnamefont{M.}~\bibnamefont{Ablikim}} \bibnamefont{et~al.}
  (\bibinfo{collaboration}{BESIII}), \bibinfo{journal}{Phys. Rev. D}
  \textbf{\bibinfo{volume}{107}}, \bibinfo{pages}{072005}
  (\bibinfo{year}{2023}{\natexlab{b}}), \eprint{2303.07629}.

\bibitem[{\citenamefont{Ablikim et~al.}(2021{\natexlab{c}})}]{BESIII:2020uqk}
\bibinfo{author}{\bibfnamefont{M.}~\bibnamefont{Ablikim}} \bibnamefont{et~al.}
  (\bibinfo{collaboration}{BESIII}), \bibinfo{journal}{Phys. Lett. B}
  \textbf{\bibinfo{volume}{814}}, \bibinfo{pages}{136110}
  (\bibinfo{year}{2021}{\natexlab{c}}), \eprint{2009.01404}.

\bibitem[{\citenamefont{Ablikim et~al.}(2024{\natexlab{a}})}]{BESIII:2023ynq}
\bibinfo{author}{\bibfnamefont{M.}~\bibnamefont{Ablikim}} \bibnamefont{et~al.}
  (\bibinfo{collaboration}{BESIII}), \bibinfo{journal}{Phys. Rev. Lett.}
  \textbf{\bibinfo{volume}{132}}, \bibinfo{pages}{081904}
  (\bibinfo{year}{2024}{\natexlab{a}}), \eprint{2307.15894}.

\bibitem[{\citenamefont{Ablikim et~al.}(2024{\natexlab{b}})}]{BESIII:2023ldb}
\bibinfo{author}{\bibfnamefont{M.}~\bibnamefont{Ablikim}} \bibnamefont{et~al.}
  (\bibinfo{collaboration}{BESIII}), \bibinfo{journal}{Phys. Rev. D}
  \textbf{\bibinfo{volume}{109}}, \bibinfo{pages}{034029}
  (\bibinfo{year}{2024}{\natexlab{b}}), \eprint{2312.12719}.

\bibitem[{\citenamefont{Gong et~al.}(2023)}]{Belle:2022dvb}
\bibinfo{author}{\bibfnamefont{G.}~\bibnamefont{Gong}} \bibnamefont{et~al.}
  (\bibinfo{collaboration}{Belle}), \bibinfo{journal}{Phys. Rev. D}
  \textbf{\bibinfo{volume}{107}}, \bibinfo{pages}{072008}
  (\bibinfo{year}{2023}), \eprint{2210.16761}.

\bibitem[{\citenamefont{Ablikim et~al.}(2022)}]{BESIII:2021rkn}
\bibinfo{author}{\bibfnamefont{M.}~\bibnamefont{Ablikim}} \bibnamefont{et~al.}
  (\bibinfo{collaboration}{BESIII}), \bibinfo{journal}{Phys. Lett. B}
  \textbf{\bibinfo{volume}{831}}, \bibinfo{pages}{137187}
  (\bibinfo{year}{2022}), \eprint{2110.04510}.

\bibitem[{\citenamefont{Ablikim et~al.}(2021{\natexlab{d}})}]{BESIII:2021aer}
\bibinfo{author}{\bibfnamefont{M.}~\bibnamefont{Ablikim}} \bibnamefont{et~al.}
  (\bibinfo{collaboration}{BESIII}), \bibinfo{journal}{Phys. Lett. B}
  \textbf{\bibinfo{volume}{820}}, \bibinfo{pages}{136557}
  (\bibinfo{year}{2021}{\natexlab{d}}), \eprint{2105.14657}.

\bibitem[{\citenamefont{Ablikim et~al.}(2021{\natexlab{e}})}]{BESIII:2020ktn}
\bibinfo{author}{\bibfnamefont{M.}~\bibnamefont{Ablikim}} \bibnamefont{et~al.}
  (\bibinfo{collaboration}{BESIII}), \bibinfo{journal}{Phys. Rev. D}
  \textbf{\bibinfo{volume}{103}}, \bibinfo{pages}{012005}
  (\bibinfo{year}{2021}{\natexlab{e}}), \eprint{2010.08320}.

\bibitem[{\citenamefont{Guo et~al.}(2025)\citenamefont{Guo, Han, Xie, and
  Chen}}]{Guo:2025rhh}
\bibinfo{author}{\bibfnamefont{B.}~\bibnamefont{Guo}},
  \bibinfo{author}{\bibfnamefont{J.}~\bibnamefont{Han}},
  \bibinfo{author}{\bibfnamefont{Y.-P.} \bibnamefont{Xie}}, \bibnamefont{and}
  \bibinfo{author}{\bibfnamefont{X.}~\bibnamefont{Chen}},
  \bibinfo{journal}{Eur. Phys. J. C} \textbf{\bibinfo{volume}{85}},
  \bibinfo{pages}{506} (\bibinfo{year}{2025}), \eprint{2501.01267}.

\bibitem[{\citenamefont{Diehl et~al.}(2003)\citenamefont{Diehl, Kroll, and
  Vogt}}]{Diehl:2002yh}
\bibinfo{author}{\bibfnamefont{M.}~\bibnamefont{Diehl}},
  \bibinfo{author}{\bibfnamefont{P.}~\bibnamefont{Kroll}}, \bibnamefont{and}
  \bibinfo{author}{\bibfnamefont{C.}~\bibnamefont{Vogt}},
  \bibinfo{journal}{Eur. Phys. J. C} \textbf{\bibinfo{volume}{26}},
  \bibinfo{pages}{567} (\bibinfo{year}{2003}), \eprint{hep-ph/0206288}.

\bibitem[{\citenamefont{Belitsky and Mueller}(2000)}]{Belitsky:2000jk}
\bibinfo{author}{\bibfnamefont{A.~V.} \bibnamefont{Belitsky}} \bibnamefont{and}
  \bibinfo{author}{\bibfnamefont{D.}~\bibnamefont{Mueller}},
  \bibinfo{journal}{Phys. Lett. B} \textbf{\bibinfo{volume}{486}},
  \bibinfo{pages}{369} (\bibinfo{year}{2000}), \eprint{hep-ph/0005028}.

\bibitem[{\citenamefont{Bertone et~al.}(2021)\citenamefont{Bertone, Dutrieux,
  Mezrag, Moutarde, and Sznajder}}]{Bertone:2021yyz}
\bibinfo{author}{\bibfnamefont{V.}~\bibnamefont{Bertone}},
  \bibinfo{author}{\bibfnamefont{H.}~\bibnamefont{Dutrieux}},
  \bibinfo{author}{\bibfnamefont{C.}~\bibnamefont{Mezrag}},
  \bibinfo{author}{\bibfnamefont{H.}~\bibnamefont{Moutarde}}, \bibnamefont{and}
  \bibinfo{author}{\bibfnamefont{P.}~\bibnamefont{Sznajder}},
  \bibinfo{journal}{Phys. Rev. D} \textbf{\bibinfo{volume}{103}},
  \bibinfo{pages}{114019} (\bibinfo{year}{2021}), \eprint{2104.03836}.

\bibitem[{\citenamefont{Liu and Ma}(2015)}]{Liu:2015jna}
\bibinfo{author}{\bibfnamefont{T.}~\bibnamefont{Liu}} \bibnamefont{and}
  \bibinfo{author}{\bibfnamefont{B.-Q.} \bibnamefont{Ma}},
  \bibinfo{journal}{Phys. Rev. D} \textbf{\bibinfo{volume}{92}},
  \bibinfo{pages}{096003} (\bibinfo{year}{2015}), \eprint{1510.07783}.

\bibitem[{\citenamefont{Afanasev et~al.}(2017)\citenamefont{Afanasev, Blunden,
  Hasell, and Raue}}]{Afanasev:2017gsk}
\bibinfo{author}{\bibfnamefont{A.}~\bibnamefont{Afanasev}},
  \bibinfo{author}{\bibfnamefont{P.~G.} \bibnamefont{Blunden}},
  \bibinfo{author}{\bibfnamefont{D.}~\bibnamefont{Hasell}}, \bibnamefont{and}
  \bibinfo{author}{\bibfnamefont{B.~A.} \bibnamefont{Raue}},
  \bibinfo{journal}{Prog. Part. Nucl. Phys.} \textbf{\bibinfo{volume}{95}},
  \bibinfo{pages}{245} (\bibinfo{year}{2017}), \eprint{1703.03874}.

\bibitem[{\citenamefont{Bytev and Tomasi-Gustafsson}(2019)}]{Bytev:2019rdc}
\bibinfo{author}{\bibfnamefont{V.~V.} \bibnamefont{Bytev}} \bibnamefont{and}
  \bibinfo{author}{\bibfnamefont{E.}~\bibnamefont{Tomasi-Gustafsson}},
  \bibinfo{journal}{Phys. Rev. C} \textbf{\bibinfo{volume}{99}},
  \bibinfo{pages}{025205} (\bibinfo{year}{2019}), \bibinfo{note}{[Erratum:
  Phys.Rev.C 106, 029902 (2022)]}, \eprint{1901.09379}.

\bibitem[{\citenamefont{Lin et~al.}(2022)\citenamefont{Lin, Hammer, and
  Mei\ss{}ner}}]{Lin:2021xrc}
\bibinfo{author}{\bibfnamefont{Y.-H.} \bibnamefont{Lin}},
  \bibinfo{author}{\bibfnamefont{H.-W.} \bibnamefont{Hammer}},
  \bibnamefont{and} \bibinfo{author}{\bibfnamefont{U.-G.}
  \bibnamefont{Mei\ss{}ner}}, \bibinfo{journal}{Phys. Rev. Lett.}
  \textbf{\bibinfo{volume}{128}}, \bibinfo{pages}{052002}
  (\bibinfo{year}{2022}), \eprint{2109.12961}.

\bibitem[{\citenamefont{McRae and Blunden}(2024)}]{McRae:2023zgu}
\bibinfo{author}{\bibfnamefont{C.}~\bibnamefont{McRae}} \bibnamefont{and}
  \bibinfo{author}{\bibfnamefont{P.~G.} \bibnamefont{Blunden}},
  \bibinfo{journal}{Phys. Rev. C} \textbf{\bibinfo{volume}{109}},
  \bibinfo{pages}{015503} (\bibinfo{year}{2024}), \eprint{2309.03892}.

\bibitem[{\citenamefont{Galynskii et~al.}(2024)\citenamefont{Galynskii, Bytev,
  and Galynsky}}]{Galynskii:2024thr}
\bibinfo{author}{\bibfnamefont{M.~V.} \bibnamefont{Galynskii}},
  \bibinfo{author}{\bibfnamefont{V.~V.} \bibnamefont{Bytev}}, \bibnamefont{and}
  \bibinfo{author}{\bibfnamefont{V.~M.} \bibnamefont{Galynsky}},
  \bibinfo{journal}{Phys. Rev. D} \textbf{\bibinfo{volume}{110}},
  \bibinfo{pages}{096017} (\bibinfo{year}{2024}), \eprint{2408.09249}.

\bibitem[{\citenamefont{Kuzmin et~al.}(2025)\citenamefont{Kuzmin, Levashko, and
  Krivoruchenko}}]{Kuzmin:2024ozz}
\bibinfo{author}{\bibfnamefont{K.~S.} \bibnamefont{Kuzmin}},
  \bibinfo{author}{\bibfnamefont{N.~M.} \bibnamefont{Levashko}},
  \bibnamefont{and} \bibinfo{author}{\bibfnamefont{M.~I.}
  \bibnamefont{Krivoruchenko}}, \bibinfo{journal}{Phys. Rev. D}
  \textbf{\bibinfo{volume}{111}}, \bibinfo{pages}{013004}
  (\bibinfo{year}{2025}), \eprint{2412.13150}.

\bibitem[{\citenamefont{Bianconi and
  Tomasi-Gustafsson}(2015)}]{Bianconi:2015owa}
\bibinfo{author}{\bibfnamefont{A.}~\bibnamefont{Bianconi}} \bibnamefont{and}
  \bibinfo{author}{\bibfnamefont{E.}~\bibnamefont{Tomasi-Gustafsson}},
  \bibinfo{journal}{Phys. Rev. Lett.} \textbf{\bibinfo{volume}{114}},
  \bibinfo{pages}{232301} (\bibinfo{year}{2015}), \eprint{1503.02140}.

\bibitem[{\citenamefont{Tomasi-Gustafsson
  et~al.}(2021)\citenamefont{Tomasi-Gustafsson, Bianconi, and
  Pacetti}}]{Tomasi-Gustafsson:2020vae}
\bibinfo{author}{\bibfnamefont{E.}~\bibnamefont{Tomasi-Gustafsson}},
  \bibinfo{author}{\bibfnamefont{A.}~\bibnamefont{Bianconi}}, \bibnamefont{and}
  \bibinfo{author}{\bibfnamefont{S.}~\bibnamefont{Pacetti}},
  \bibinfo{journal}{Phys. Rev. C} \textbf{\bibinfo{volume}{103}},
  \bibinfo{pages}{035203} (\bibinfo{year}{2021}), \eprint{2012.14656}.

\bibitem[{\citenamefont{Lomon and Pacetti}(2012)}]{Lomon:2012pn}
\bibinfo{author}{\bibfnamefont{E.~L.} \bibnamefont{Lomon}} \bibnamefont{and}
  \bibinfo{author}{\bibfnamefont{S.}~\bibnamefont{Pacetti}},
  \bibinfo{journal}{Phys. Rev. D} \textbf{\bibinfo{volume}{85}},
  \bibinfo{pages}{113004} (\bibinfo{year}{2012}), \bibinfo{note}{[Erratum:
  Phys.Rev.D 86, 039901 (2012)]}, \eprint{1201.6126}.

\bibitem[{\citenamefont{Qian et~al.}(2023)\citenamefont{Qian, Liu, Cao, and
  Liu}}]{Qian:2022whn}
\bibinfo{author}{\bibfnamefont{R.-Q.} \bibnamefont{Qian}},
  \bibinfo{author}{\bibfnamefont{Z.-W.} \bibnamefont{Liu}},
  \bibinfo{author}{\bibfnamefont{X.}~\bibnamefont{Cao}}, \bibnamefont{and}
  \bibinfo{author}{\bibfnamefont{X.}~\bibnamefont{Liu}},
  \bibinfo{journal}{Phys. Rev. D} \textbf{\bibinfo{volume}{107}},
  \bibinfo{pages}{L091502} (\bibinfo{year}{2023}), \eprint{2211.11555}.

\bibitem[{\citenamefont{Yan et~al.}(2024)\citenamefont{Yan, Chen, Li, and
  Xie}}]{Yan:2023nlb}
\bibinfo{author}{\bibfnamefont{B.}~\bibnamefont{Yan}},
  \bibinfo{author}{\bibfnamefont{C.}~\bibnamefont{Chen}},
  \bibinfo{author}{\bibfnamefont{X.}~\bibnamefont{Li}}, \bibnamefont{and}
  \bibinfo{author}{\bibfnamefont{J.-J.} \bibnamefont{Xie}},
  \bibinfo{journal}{Phys. Rev. D} \textbf{\bibinfo{volume}{109}},
  \bibinfo{pages}{036033} (\bibinfo{year}{2024}), \eprint{2312.04866}.

\bibitem[{\citenamefont{Huang and Ferroli}(2021)}]{Huang:2021xte}
\bibinfo{author}{\bibfnamefont{G.}~\bibnamefont{Huang}} \bibnamefont{and}
  \bibinfo{author}{\bibfnamefont{R.~B.} \bibnamefont{Ferroli}}
  (\bibinfo{collaboration}{BESIII}), \bibinfo{journal}{Natl. Sci. Rev.}
  \textbf{\bibinfo{volume}{8}}, \bibinfo{pages}{nwab187}
  (\bibinfo{year}{2021}), \eprint{2111.08425}.

\bibitem[{\citenamefont{Yang et~al.}(2024)\citenamefont{Yang, Guo, Li, Dai,
  Haidenbauer, and Mei\ss{}ner}}]{Yang:2024iuc}
\bibinfo{author}{\bibfnamefont{Q.-H.} \bibnamefont{Yang}},
  \bibinfo{author}{\bibfnamefont{D.}~\bibnamefont{Guo}},
  \bibinfo{author}{\bibfnamefont{M.-Y.} \bibnamefont{Li}},
  \bibinfo{author}{\bibfnamefont{L.-Y.} \bibnamefont{Dai}},
  \bibinfo{author}{\bibfnamefont{J.}~\bibnamefont{Haidenbauer}},
  \bibnamefont{and} \bibinfo{author}{\bibfnamefont{U.-G.}
  \bibnamefont{Mei\ss{}ner}}, \bibinfo{journal}{JHEP}
  \textbf{\bibinfo{volume}{08}}, \bibinfo{pages}{208} (\bibinfo{year}{2024}),
  \eprint{2404.12448}.

\bibitem[{\citenamefont{Cao et~al.}(2022)\citenamefont{Cao, Dai, and
  Lenske}}]{Cao:2021asd}
\bibinfo{author}{\bibfnamefont{X.}~\bibnamefont{Cao}},
  \bibinfo{author}{\bibfnamefont{J.-P.} \bibnamefont{Dai}}, \bibnamefont{and}
  \bibinfo{author}{\bibfnamefont{H.}~\bibnamefont{Lenske}},
  \bibinfo{journal}{Phys. Rev. D} \textbf{\bibinfo{volume}{105}},
  \bibinfo{pages}{L071503} (\bibinfo{year}{2022}), \eprint{2109.15132}.

\bibitem[{\citenamefont{Alexandrou et~al.}(2025)\citenamefont{Alexandrou,
  Bacchio, Finkenrath, Iona, Koutsou, Li, and Spanoudes}}]{Alexandrou:2024ozj}
\bibinfo{author}{\bibfnamefont{C.}~\bibnamefont{Alexandrou}},
  \bibinfo{author}{\bibfnamefont{S.}~\bibnamefont{Bacchio}},
  \bibinfo{author}{\bibfnamefont{J.}~\bibnamefont{Finkenrath}},
  \bibinfo{author}{\bibfnamefont{C.}~\bibnamefont{Iona}},
  \bibinfo{author}{\bibfnamefont{G.}~\bibnamefont{Koutsou}},
  \bibinfo{author}{\bibfnamefont{Y.}~\bibnamefont{Li}}, \bibnamefont{and}
  \bibinfo{author}{\bibfnamefont{G.}~\bibnamefont{Spanoudes}},
  \bibinfo{journal}{Phys. Rev. D} \textbf{\bibinfo{volume}{111}},
  \bibinfo{pages}{054505} (\bibinfo{year}{2025}), \eprint{2412.01535}.

\end{thebibliography}

\end{document}